\documentclass[journal]{IEEEtran}

\usepackage[utf8]{inputenc}
\usepackage{glossaries}
\usepackage[symbols,nogroupskip]{glossaries-extra}
\usepackage{bm}
\usepackage{amssymb}
\usepackage{amsmath}
\usepackage{amsfonts}
\usepackage{mathrsfs}
\usepackage{bbm}
\usepackage{scalerel} 
\usepackage{graphicx}
\usepackage{subfigure}
\usepackage{cite}
\usepackage{enumerate}
\usepackage{url}
\usepackage{epstopdf}
\usepackage{algorithm}
\usepackage[noend]{algpseudocode}
\usepackage{algorithmicx,algorithm}
\usepackage{float}
\usepackage{subfigure}
\usepackage{amsthm}
\usepackage{color}
\usepackage{enumitem}
\usepackage{clipboard}
\usepackage[normalem]{ulem} 
\usepackage{framed}

\usepackage[colorinlistoftodos,prependcaption,textsize=scriptsize]{todonotes}
\usepackage{xargs} 
\usepackage{cooltooltips}
\usepackage{xcolor}

\newcommand{\beq}{\begin{equation}}
\newcommand{\eeq}{\end{equation}}
\newcommand{\beqq}{\begin{equation*}}
\newcommand{\eeqq}{\end{equation*}}
\newcommand{\mbeqa}{\begin{align}}
\newcommand{\meeqa}{\end{align}}

\makeatletter
\def\@eqnnum{{\normalfont \normalcolor \theequation}}
\makeatother

\makeatletter
\newcommand{\manuallabel}[2]{\def\@currentlabel{#2}\label{#1}}
\makeatother

\makeatletter
\newcommand{\customlabel}[2]{%
   \protected@write \@auxout {}{\string \newlabel {#1}{{#2}{\thepage}{#2}{#1}{}} }%
   \hypertarget{#1}{}
}
\makeatother

\newtheorem{proposition}{\bf Proposition}

\ifodd 0
\newcommand{\revise}{\textcolor[rgb]{0,0,0}}
\newcommand{\revises}{\textcolor[rgb]{0.1,0.2,0.6}}

\else
\newcommand{\revise}{}
\newcommand{\revises}{}

\fi 

\newif\ifToShow
\ToShowfalse
\ifToShow

\newcommand{\red}{\textcolor[rgb]{0,0,0}}

\newcommand{\frev}{\textcolor[rgb]{0,0,0.7}}
\newcommand{\srev}{\textcolor[rgb]{0,0,0.7}}
\else

\newcommand{\red}{\textcolor[rgb]{0,0,0}}

\newcommand{\frev}{\textcolor[rgb]{0,0,0}}
\newcommand{\srev}{\textcolor[rgb]{0,0,0}}
\fi

\allowdisplaybreaks

\usepackage{clipboard}

\newcommand{\hil}{\mathrm{H}}
\newcommand{\tran}{\top}
\newcommand{\iu}{\mathrm{i}}

\newcommand{\trace}{\operatorname{Tr}}

\newcommand{\diag}{\operatorname{diag}}
\newcommand{\dd}{\operatorname{d}}
\renewcommand{\trace}{\operatorname{tr}}
\renewcommand{\diag}{\bm{\mathrm{diag}}}

\newcommand{\prox}{\bm{\mathrm{prox}}}
\newcommand{\outter}{{\mathrm{out}}} 
\newcommand{\inner}{{\mathrm{in}}} 
\newcommand{\rx}{{\mathrm{Rx}}}

\newcommand{\aod}{\mathrm{AoD}}
\newcommand{\aoa}{\mathrm{AoA}}
\newcommand{\los}{\mathrm{LoS}}
\newcommand{\mulpath}{\mathrm{MP}}
\newcommand{\user}{\mathrm{U}}
\newcommand{\elem}{\mathrm{E}}

\DeclareRobustCommand
\cpdots{\mathinner{\mkern-2mu ...\mkern-2mu}}

\makeglossaries
\newacronym{rhs}{RHS}{reconfigurable holographic surface}

\newacronym{isac}{ISAC}{integrated sensing and communications}

\newacronym{crlb}{CRLB}{Cramer-Rao lower bound}

\newacronym{sgd}{SGD}{stochastic gradient descent}

\newacronym{ftl}{FTL}{federated transfer learning}
\newacronym{fl}{FL}{federated learning}

\newacronym{ml}{ML}{machine learning}

\newacronym{roi}{ROI}{region of interest}

\newacronym{pos_est}{position estimator}{position estimator}

\newacronym{bs}{BS}{base station}

\newacronym{rf}{RF}{radio frequency}

\newacronym{em}{EM}{electromagnetic}

\newacronym{ofdm}{OFDM}{orthogonal frequency division multiplexing}

\newacronym{pos_process}{DPP}{positioning process}

\newacronym{pos_protocol}{DPP}{positioning protocol}

\newacronym{pos_acc}{PP}{positioning precision}

\newacronym{mmht}{MMT}{multi-band multi-pattern transmission}

\newacronym{me}{meta-element}{metamaterial element}

\newacronym{aod}{AoD}{angle of departure}
\newacronym{aoa}{AoA}{angle of arrival}

\newacronym{los}{LoS}{line-of-sight}

\newacronym{wss}{WSS}{wide-sense stationary}

\newacronym{mse}{MSE}{mean squared error}

\newacronym{rss}{RSS}{received signal strength}

\newacronym{bcd}{BCD}{block coordinate descent}

\newacronym{gps}{GPS}{Global Positioning System}

\newacronym{uav}{UAV}{unmanned aerial vehicle}

\newacronym{psgd}{PSGD}{proximal stochastic gradient descent}

\newacronym{rfid}{RFID}{radio frequency identification}

\newacronym{snr}{SNR}{signal-to-noise ratio}

\newacronym{ris}{RIS}{reconfigurable intelligent surface}

\newacronym{mb}{MB}{multi-band}

\newacronym{tnn}{TNN}{transformer neural network}

\newacronym{dna}{DA}{digital and analog}

\newacronym{cdf}{CDF}{cumulative distribution function}
\newacronym{pdf}{PDF}{probability density function}

\newacronym{fd}{FD}{finite difference}

\newacronym{fim}{FIM}{Fisher information matrix}
\newacronym{dp}{DP}{differentially private}

\newacronym{mlp}{MLP}{multilayer perceptron}

\newacronym{rms}{RMS}{root mean square}
\newglossaryentry{SysName}
{
    name={HoloFed},
    description={the name of the proposed system}
}

\newglossaryentry{MetaElem}
{
    name={metamaterial element},
    description={Metamaterial element of RHS}
}

\newglossaryentry{HoloMat}
{
    name={holographic pattern matrix},
    description={holographic pattern matrix of RHS}
}

\newglossaryentry{tarSpace}
{
	name={\ensuremath{\mathcal P}},
	sort={p},
	description={region of interest}
}

\newglossaryentry{numUser}
{
	name={\ensuremath{U}},
	sort={p},
	description={number of users}
}

\newglossaryentry{numElem}
{
	name={\ensuremath{N_{\mathrm{E}}}},
	sort={p},
	description={number of metamaterial elements}
}

\newglossaryentry{numFeed}
{
    name={\ensuremath{K}},
    sort={k},
    description={number of feeds in RHS}
}

\newglossaryentry{numState}
{
    name={\ensuremath{N_{\mathrm{S}}}},
    sort={n},
    description={number of configurable states of each meta element}
}

\newglossaryentry{setRadCoeff}
{
    name={\ensuremath{\mathcal R}},
    sort={r},
    description={set of radiation coefficients}
}

\newglossaryentry{numBand}
{
	name={\ensuremath{N_{\mathrm{B}}}},
	sort={n},
	description={number of downlink ofdm bands}
}

\newglossaryentry{numSBand}
{
	name={\ensuremath{N_{\mathrm{SB}}}},
	sort={p},
	description={number of sub-band of each ofdm band}
}

\newglossaryentry{ulFreq}
{
	name={\ensuremath{f_{\mathrm{c}}^{\mathrm{u}}}},
	sort={p},
	description={carrier frequency of UL transmission}
}

\newglossaryentry{ulBandwid}
{
	name={\ensuremath{B^{\mathrm{u}}}},
	sort={p},
	description={bandwidth of UL transmission}
}

\newglossaryentry{userMaxSpeed}
{
	name={\ensuremath{v_{\max}}},
	sort={p},
	description={user max speed}
}

\newglossaryentry{userPos}
{
	name={\ensuremath{\bm p}},
	sort={p},
	description={user position}
}

\newglossaryentry{userPos_n}
{
	name={\ensuremath{\bm p_n}},
	sort={p},
	description={user $n$ position}
}

\newglossaryentry{userPos_n^t}
{
	name={\ensuremath{\bm p_n^{(t)}}},
	sort={p},
	description={user $n$ position in process $t$}
}

\newglossaryentry{userPos_est}
{
	name={\ensuremath{\tilde{\bm p}}},
	sort={p},
	description={estimated user position}
}

\newglossaryentry{userPos_est_n^t}
{
	name={\ensuremath{\tilde{\bm p}_n^{(t)}}},
	sort={p},
	description={estimated user position}
}

\newglossaryentry{userDistribu}
{
	name={\ensuremath{\varGamma^{\user}}},
	sort={p},
	description={distribution of user's position}
}

\newglossaryentry{posFunc}
{
    name={\ensuremath{\bm f}},
    sort={p},
    description={positioning estimator of users}
}

\newglossaryentry{posFunc_w}
{
    name={\ensuremath{\bm f^{\bm w}}},
    sort={p},
    description={positioning estimator of users}
}

\newglossaryentry{numFrame}
{
    name={\ensuremath{F}},
    sort={k},
    description={number of frames sent in each band}
}

\newglossaryentry{sigMat_i}
{
    name={\ensuremath{\bm S_i}},
    sort={s},
    description={signal matrix of the $i$-th band}
}

\newglossaryentry{sigMat_ij}
{
    name={\ensuremath{\bm S_{i,j}}},
    sort={s},
    description={signal matrix of the $i$-th band}
}

\newglossaryentry{sigSet}
{
    name={\ensuremath{\{\bm S_{i,j}\}_{i,j}}},
    sort={s},
    description={set of signal matrices}
}

\newglossaryentry{sigSet_opt}
{
    name={\ensuremath{\{\bm S^*_{i,j}\}_{i,j}}},
    sort={s},
    description={set of signal matrices}
}

\newglossaryentry{sigVec_ijq}
{
    name={\ensuremath{\bm s_{i,j}^{(q)}}},
    sort={s},
    description={feed signal vector in the $i$-th band, $j$-th sub-band, and $q$-th frame}
}

\newglossaryentry{codeMat_i}
{
    name={\ensuremath{\bm C_i}},
    sort={c},
    description={code matrix of the RHS in the $i$-th band}
}

\newglossaryentry{codeSet}
{
    name={\ensuremath{\{\gls{codeMat_i}\}_i}},
    sort={c},
    description={set of code matrix of RHS}
}

\newglossaryentry{codeSet_opt}
{
    name={\ensuremath{\{\bm C_i^*\}_i}},
    sort={c},
    description={set of code matrix of RHS}
}

\newglossaryentry{codeVec_iq}
{
    name={\ensuremath{\bm c_{i}^{(q)}}},
    sort={s},
    description={code vector of RHS in the $i$-th band and $q$-th frame}
}

\newglossaryentry{code_iqm}
{
    name={\ensuremath{c_{i,m}^{(q)}}},
    sort={s},
    description={code of Meta-elemt $m$ in the $i$-th band and $q$-th frame}
}

\newglossaryentry{recvSigMat}
{
    name={\ensuremath{\bm Y_{\mathrm{Rx}}}},
    sort={y},
    description={received signals arranged as a matrix in the all the bands}
}

\newglossaryentry{recvSigMat_n^t}
{
    name={\ensuremath{\bm Y_{\mathrm{Rx},n}^{(t)}}},
    sort={y},
    description={received signals arranged as a matrix in the all the bands for user $n$ and positioning process $t$}
}

\newglossaryentry{recvSigVec_i}
{
    name={\ensuremath{\bm y_{i}}},
    sort={y},
    description={received signals arranged as a vector in the Band $i$}
}

\newglossaryentry{recvSigSet}
{
    name={\ensuremath{\mathcal Y}},
    sort={y},
    description={set of all the received signal matrices}
}

\newglossaryentry{posFeed_k}
{
    name={\ensuremath{\bm p^{\mathrm{F}}_k}},
    sort={p},
    description={position of Feed $k$}
}

\newglossaryentry{posElem_m}
{
    name={\ensuremath{\bm p^{\mathrm{E}}_m}},
    sort={p},
    description={position of Meta-elem $m$}
}

\newglossaryentry{posElem_1}
{
    name={\ensuremath{\bm p^{\mathrm{E}}_1}},
    sort={p},
    description={position of Meta-elem $1$}
}

\newglossaryentry{holoVec_ijqm}
{
    name={\ensuremath{\tau_{i,j,m}^{(q)}}},
    sort={t},
    description={holograph record/radiation pattern in Frame $q$ of Band $i$ Sub-band $j$}
}

\newglossaryentry{recvSig_ijq}
{
    name={\ensuremath{y_{i,j}^{(q)}}},
    sort={y},
    description={received signal of user}
}

\newglossaryentry{noise_ijq}
{
	name={\ensuremath{e_{i,j}^{(q)}}},
	sort={p},
	description={thermal noise signal}
}

\newglossaryentry{losgain_ijm}
{
    name={\ensuremath{h^{\los}_{i,j,m}}},
    sort={h},
    description={LoS channel gain}
}

\newglossaryentry{mpgain_ijmq}
{
    name={\ensuremath{h^{\mulpath,(q)}_{i,j,m}}},
    sort={h},
    description={multipath channel gain}
}

\newglossaryentry{mpgainVec_ijq}
{
    name={\ensuremath{\bm h^{\mulpath,(q)}_{i,j}}},
    sort={h},
    description={multipath channel gain vector}
}

\newglossaryentry{noisepower_i}
{
    name={\ensuremath{\sigma_i^2}},
    sort={s},
    description={noise power}
}

\newglossaryentry{covmat_i}
{
    name={\ensuremath{\bm V_i}},
    sort={v},
    description={covariance matrix of multipath gain}
}

\newglossaryentry{angularResponse_i}
{
    name={\ensuremath{\bm \alpha_i}},
    sort={a},
    description={angular array response function of RHS in Band $i$}
}

\newglossaryentry{covCoefFuncFreq_i}
{
	name={\ensuremath{\rho_{\mathrm{f},i}}},
	sort={r},
	description={covariance coefficient function in the frequency domain}
}

\newglossaryentry{covCoefFuncTime_i}
{
	name={\ensuremath{\rho_{\mathrm{t},i}}},
	sort={r},
	description={covariance coefficient function in the time domain}
}

\newglossaryentry{rmsValue_i}
{
    name={\ensuremath{\sigma_{\mathrm{rms},i}}},
    sort={s},
    description={RMS delay spread value}
}

\newglossaryentry{dopplerFreq_i}
{
    name={\ensuremath{f_{\mathrm{D},i}}},
    sort={f},
    description={Doppler frequency}
}

\newglossaryentry{holoMat_i}
{
    name={\ensuremath{\bm T_i}},
    sort={t},
    description={holographic radiation pattern of RHS}
}

\newglossaryentry{bPropMat_i}
{
    name={\ensuremath{\bm B_{i}}},
    sort={b},
    description={on-board propagation coefficient matrix}
}

\newglossaryentry{gainMat_i}
{
    name={\ensuremath{\bm G_i}},
    sort={g},
    description={gain matrix in Band $i$}
}

\newglossaryentry{valueSet_i}
{
    name={\ensuremath{\{1,\cpdots,\gls{numBand}\}}},
    sort={0},
    description={shortcut forall $i$}
}

\newglossaryentry{valueSet_j}
{
    name={\ensuremath{\{1,\cpdots,\gls{numSBand}\}}},
    sort={0},
    description={shortcut forall $j$}
}

\newglossaryentry{valueSet_k}
{
    name={\ensuremath{\{1,\cpdots,\gls{numFeed}\}}},
    sort={0},
    description={shortcut forall $k$}
}

\newglossaryentry{valueSet_q}
{
    name={\ensuremath{\{1,\cpdots,\gls{numFrame}\}}},
    sort={0},
    description={shortcut forall $q$}
}

\newglossaryentry{valueSet_m}
{
    name={\ensuremath{\{1,\cpdots,\gls{numElem}\}}},
    sort={0},
    description={shortcut forall $m$}
}

\newglossaryentry{valueSet_n}
{
    name={\ensuremath{\{1,\cpdots,\gls{numUser}\}}},
    sort={0},
    description={shortcut forall $n$}
}

\newglossaryentry{valueSet_t}
{
    name={\ensuremath{\{0,\cpdots, T\}}},
    sort={0},
    description={shortcut forall $t$}
}

\newglossaryentry{vecUserSpeed}
{
    name={\ensuremath{\bm v^{\user}}},
    sort={0},
    description={velocity vector of user}
}

\newglossaryentry{durPosProc}
{
    name={\ensuremath{D_{\mathrm{pos}}}},
    sort={0},
    description={duration of positioning process}
}

\newglossaryentry{crlb_userpos}
{
    name={\ensuremath{\mathrm{CRLB}(\gls{userPos})}},
    sort={0},
    description={CRLB at certain user position}
}

\newglossaryentry{fimMat_userpos}
{
    name={\ensuremath{\bm I_{\mathrm{FIM}}(\gls{userPos})}},
    sort={0},
    description={FIM at certain user position}
}

\newglossaryentry{fimMat_userpos_Sij}
{
    name={\ensuremath{\bm I_{\mathrm{FIM}}(\gls{userPos}; \gls{sigMat_ij})}},
    sort={0},
    description={FIM at certain user position}
}

\newglossaryentry{invfimMat_userpos}
{
    name={\ensuremath{\bm I^{-1}_{\mathrm{FIM}}(\gls{userPos})}},
    sort={0},
    description={FIM at certain user position}
}

\newglossaryentry{reshapeFunc}
{
    name={\ensuremath{\bm R}},
    sort={l},
    description={reshape function}
}
\newglossaryentry{setSamples}
{
    name={\ensuremath{\mathcal S_{\mathrm{sam}}}},
    sort={l},
    description={side length of ROI in z-axis}
}
\newglossaryentry{inv2fimMat_userpos}
{
    name={\ensuremath{\bm I^{-2}_{\mathrm{FIM}}(\gls{userPos})}},
    sort={0},
    description={FIM at certain user position}
}
\newglossaryentry{inv2fimMat_userpos_Sij}
{
    name={\ensuremath{\bm I^{-2}_{\mathrm{FIM}}(\gls{userPos}; \gls{sigMat_ij})}},
    sort={0},
    description={FIM at certain user position}
}
\newglossaryentry{numSamples}
{
    name={\ensuremath{N_{\mathrm{sam}}}},
    sort={l},
    description={side length of ROI in z-axis}
}
\newglossaryentry{userMaxPower}
{
    name={\ensuremath{P_{\max}}},
    sort={s},
    description={user's max power}
}
\newglossaryentry{numMaxIteration}
{
    name={\ensuremath{N_{\mathrm{itr}}}},
    sort={l},
    description={maximum number of iteration}
}

\newglossaryentry{privacyCoeffVec}
{
    name={\ensuremath{\bm \varepsilon_{\mathrm{DP}}}},
    sort={l},
    description={privacy demand vector}
}

\newglossaryentry{roi_sideLen_x}
{
    name={\ensuremath{l_{\mathrm x}}},
    sort={l},
    description={side length of ROI in x-axis}
}

\newglossaryentry{roi_sideLen_y}
{
    name={\ensuremath{l_{\mathrm y}}},
    sort={l},
    description={side length of ROI in y-axis}
}

\newglossaryentry{roi_sideLen_z}
{
    name={\ensuremath{l_{\mathrm z}}},
    sort={l},
    description={side length of ROI in z-axis}
}

\newglossaryentry{subBandwidth}
{
    name={\ensuremath{W}},
    sort={p},
    description={sub-band bandwidth}
}

\newglossaryentry{durFrame}
{
    name={\ensuremath{\Delta_{\mathrm{t}}}},
    sort={0},
    description={duration of each frame}
}

\newglossaryentry{noisePowerDensity}
{
    name={\ensuremath{P_{\mathrm{N}}}},
    sort={p},
    description={spectral power density}
}

\newglossaryentry{noisePower}
{
    name={\ensuremath{\sigma^2}},
    sort={s},
    description={noise power}
}

\newglossaryentry{numUpdate}
{
    name={\ensuremath{N_{\mathrm{us}}}},
    sort={l},
    description={number of update steps}
}

\newglossaryentry{simulDomData}
{
    name={\ensuremath{\mathcal D_{\mathrm{src}}}},
    sort={l},
    description={source domain dataset}
}

\newglossaryentry{numSimulDomData}
{
    name={\ensuremath{N_{\mathrm{src}}}},
    sort={l},
    description={source/empirical domain dataset}
}

\newglossaryentry{ftlDomData}
{
    name={\ensuremath{\mathcal D_{\mathrm{FTL}}}},
    sort={l},
    description={collective data set for FTL}
}

\newglossaryentry{netwCoeff}
{
    name={\ensuremath{\bm w}},
    sort={l},
    description={neural network coefficient}
}

\newglossaryentry{sizeCoeff}
{
    name={\ensuremath{B_{\mathrm{para}}}},
    sort={l},
    description={data size of parameter vector}
}

\newglossaryentry{numHotSpot}
{
    name={\ensuremath{N_{\mathrm{hs}}}},
    sort={l},
    description={number of hot spots}
}

\newglossaryentry{numFTLepo}
{
    name={\ensuremath{N_{\mathrm{epo}}}},
    sort={l},
    description={number of epoch}
}

\newglossaryentry{dpEpsilon_n}
{
    name={\ensuremath{\epsilon_{\mathrm{dp}, n}}},
    sort={l},
    description={DP epsilon coefficient}
}

\newglossaryentry{dpEpsilon_n2}
{
    name={\ensuremath{\epsilon_{\mathrm{dp}, n}^2}},
    sort={l},
    description={DP epsilon coefficient}
}

\newglossaryentry{dpEpsilonVec}
{
    name={\ensuremath{\bm \epsilon_{\mathrm{dp}}}},
    sort={l},
    description={DP epsilon parameter vector}
}

\newglossaryentry{dpDelta}
{
    name={\ensuremath{\delta_{\mathrm{dp}}}},
    sort={l},
    description={DP delta coefficient}
}

\newglossaryentry{dpSigma_n}
{
    name={\ensuremath{\sigma_{\mathrm{dp}, n}}},
    sort={l},
    description={DP sigma coefficient}
}

\newglossaryentry{dpSigma_n2}
{
    name={\ensuremath{\sigma_{\mathrm{dp}, n}^2}},
    sort={l},
    description={DP sigma 2 coefficient}
}

\newglossaryentry{direPatternFeed}
{
    name={\ensuremath{g_{i,j}^{\mathrm{F}}}},
    sort={l},
    description={direction pattern of feed}
}

\newglossaryentry{direPatternElem}
{
    name={\ensuremath{g_{i,j}^{\mathrm{E}}}},
    sort={l},
    description={direction pattern of element}
}

\newglossaryentry{direPatternUser}
{
    name={\ensuremath{g_{i,j}^{\mathrm{U}}}},
    sort={l},
    description={direction pattern of element}
}

\newglossaryentry{dimCoeff}
{
    name={\ensuremath{N_{\mathrm{para}}}},
    sort={l},
    description={direction pattern of element}
}

\newglossaryentry{dimCoeff_reg}
{
    name={\ensuremath{N_{\mathrm{reg}}}},
    sort={l},
    description={direction pattern of element}
}

\newglossaryentry{dimCoeff_feat}
{
    name={\ensuremath{N_{\mathrm{feat}}}},
    sort={l},
    description={direction pattern of element}
}

\newglossaryentry{lrVec}
{
    name={\ensuremath{\bm \eta}},
    sort={l},
    description={learning rate vector}
}

\newglossaryentry{numPP}
{
	name={\ensuremath{T}},
	sort={p},
	description={number of positioning process}
}

\begin{document}

\title{{HoloFed: Environment-Adaptive Positioning via Multi-band Reconfigurable Holographic Surfaces and Federated Learning}}

\author{
\IEEEauthorblockN{
{Jingzhi~Hu},~\IEEEmembership{Member,~IEEE},
{Zhe~Chen},~\IEEEmembership{Member,~IEEE},
{Tianyue~Zheng},~\IEEEmembership{Graduate~Student~Member,~IEEE},
{Robert~Schober},~\IEEEmembership{Fellow,~IEEE},
and~{Jun~Luo},~\IEEEmembership{Senior~Member,~IEEE}
}
\thanks{© 2023 IEEE.  Personal use of this material is permitted.  Permission from IEEE must be obtained for all other uses, in any current or future media, including reprinting/republishing this material for advertising or promotional purposes, creating new collective works, for resale or redistribution to servers or lists, or reuse of any copyrighted component of this work in other works.}
\thanks{This research is supported by National Research Foundation, Singapore and Infocomm Media Development Authority under its Future Communications Research \& Development Programme grant FCP-NTU-RG-2022-015, and in part by the German Research Foundation (DFG) under project SFB 1483 (Project-ID 442419336 Empkins) and the BMBF under the program of ``Souverän. Digital. Vernetzt.'' joint project 6G-RIC (Project-ID 16KISK023). \emph{(Corresponding author: Zhe Chen.)}}
\thanks{This work has been presented in part at IEEE ICC 2023~\cite{Hu2023ICC}. }
\thanks{
 J. Hu, T. Zheng, and J. Luo are with School of Computer Science and Engineering, Nanyang Technological University, Singapore 639798, Singapore~(email: jingzhi.hu@ntu.edu.sg, tianyue002@ntu.edu.sg, junluo@ntu.edu.sg).}
 \thanks{Z. Chen is with Intelligent Networking and Computing Research Center and School of Computer Science, Fudan University, Shanghai 200438, China~(email: zhechen@fudan.edu.cn).}
 \thanks{R. Schober is with the Institute of Digital Communications, Friedrich–Alexander University of Erlangen–Nuremberg, 91058 Erlangen, Germany~(e-mail: robert.schober@fau.de).}
}

\maketitle
\begin{abstract}
Positioning is an essential service for various applications and is expected to be integrated with existing communication infrastructures in 5G and 6G. Though current Wi-Fi and cellular base stations (BSs) can be used to support this integration, the resulting precision is unsatisfactory due to the lack of precise control of the wireless signals. Recently, BSs adopting reconfigurable holographic surfaces (RHSs) have been advocated for positioning as RHSs' large number of antenna elements enable generation of arbitrary and highly-focused signal beam patterns. However, existing designs face two major challenges: i) RHSs only have limited operating bandwidth, and ii) the positioning methods cannot adapt to the diverse environments encountered in practice. To overcome these challenges, we present HoloFed, a system providing high-precision environment-adaptive user positioning services by exploiting \emph{multi-band} (MB)-RHS and \emph{federated learning} (FL). For improving the positioning performance, a lower bound on the error variance is obtained and utilized for guiding MB-RHS's digital and analog beamforming design. For better adaptability while preserving privacy, an FL framework is proposed for users to collaboratively train a position estimator, where we exploit the transfer learning technique to handle the lack of position labels of the users. Moreover, a scheduling algorithm for the BS to select which users train the position estimator is designed, jointly considering the convergence and efficiency of FL. Our performance evaluation based on simulations confirms that HoloFed achieves a $57$\% lower positioning error variance compared to a beam-scanning baseline and can effectively adapt to diverse environments.

\end{abstract}
\begin{IEEEkeywords}
Positioning, reconfigurable holographic surfaces, beamforming, federated learning.
\end{IEEEkeywords}

\section[Introduction]{Introduction}
\label{sec: intro}

In 5G and 6G wireless systems, positioning is an essential service fundamental to both user location awareness and improved communication~\cite{Dardari2022LOS, elzanaty2021towards}, and thus has an ever-expanding range of applications in civil and military scenarios~\cite{Win2018ATheoretical}. 
Among all the available positioning techniques, the satellite-based \acrfull{gps} is the most widely used one and can achieve high precision in ideal outdoor environments.
Nevertheless, it has the drawbacks of consuming a lot of energy and frequently loosing track when buildings block the satellite signals~\cite{wang2018learning,cherian2018parkloc}.
The loss of \acrshort{gps} signals can take place in various outdoor, indoor, and underground scenarios, creating many \acrshort{gps}-deprived regions where users receive poor positioning services.

\revise{To provide positioning services in \acrshort{gps}-deprived regions, many different GPS-free alternatives have been studied, including video-based~\cite{Morar2020Comprehensive}, radar-based~\cite{Zhou2020MMW}, and \acrfull{rfid}-based~\cite{Motroni2021ASurvey} positioning techniques.}
Though the above-mentioned techniques can achieve high precision, they all require additional infrastructure which can be cost-prohibitive for realizing ubiquitous positioning~\cite{cherian2018parkloc}.
To reduce the infrastructure cost, \acrfull{isac} has been proposed as a key enabling technology for 6G, integrating sensing and positioning functions into the existing communication infrastructures, e.g., Wi-Fi and cellular base stations~(\acrshort{bs}s).
Nevertheless, such piggyback positioning systems generally cannot ensure high positioning precision, mainly due to their limited bandwidth and comparatively low number of antenna elements.

Thanks to the recent development of metamaterial-based reconfigurable holographic surfaces (\acrshort{rhs}s), one may achieve a cost-efficient increase in communication rate, while potentially enhancing the precision of the piggyback positioning services at the same time.
This potential is largely attributed to \acrshort{rhs}s' characteristic of comprising a massive number of metamaterial antenna elements (\emph{\acrshort{me}s} for short), which are densely arranged and have much smaller spatial spacing than half of their operating wavelength.
Such a dense arrangement enables \acrshort{rhs}s to synthesize arbitrary wavefronts and beam shapes~\cite{Zhang2022Holographic}, suggesting their strong capability in manipulating \acrfull{em} waves~\cite{Zhang2022Holographic}.
Utilizing this capability, \acrshort{bs}s equipped with an \acrshort{rhs} can focus transmitted signals into sharp beams to enhance the \acrfull{snr} of users and to probe the region of interest with high resolution and precision.

A few \acrshort{rhs}-based systems have been proposed for positioning in the literature~\cite{ZhangX2022Holographic,Zhang2022Holographic}: an \acrshort{rhs}-based \acrshort{isac} system is proposed to generate signal beams for both sensing and communication with high gains in~\cite{Zhang2022Holographic}, and an \acrshort{rhs} is leveraged for target detection with high accuracy yet at low power and cost in~\cite{ZhangX2022Holographic}.
In this context, it is also worthwhile to mention positioning methods exploiting reconfigurable intelligent surfaces~(\acrshort{ris}s)~\cite{zhang2020towards,Elzanaty2021Reconfigurable, Zohair2021Near,Wang2022Location,Dardari2022LOS,bjornson2022reconfigurable,nguyen2020reconfigurable} due to their intrinsic similarity.
\revises{Although \acrshort{ris}s differ from \acrshort{rhs}s in i) the signal feeding scheme (\acrshort{ris}s' over-the-air propagation as opposed to \acrshort{rhs}s' on-board propagation) and thus ii) take up a larger space, 
the positioning methods for \acrshort{ris}- and \acrshort{rhs}-based systems are largely comparable since they both leverage massive numbers of \acrshort{me}s for analog beamforming. }
In~\cite{zhang2020towards}, the authors utilize an \acrshort{ris} to generate distinguishable signals at different positions and employ a positioning method based on maximum likelihood estimation~(MLE).
The authors of \cite{Elzanaty2021Reconfigurable, Zohair2021Near} also employ MLE-based positioning methods, and they optimize the beamforming of the \acrshort{ris} by minimizing the \acrfull{crlb} on the positioning error.
The beamforming optimization problem is then extended to scenarios involving multiple \acrshort{ris}s and obstacles in~\cite{Wang2022Location}.
In addition to the MLE-based methods, the authors of \cite{Dardari2022LOS,bjornson2022reconfigurable} propose positioning methods based on estimating the time differences among the signals arriving from an \acrshort{ris}.
Moreover, the authors of \cite{nguyen2020reconfigurable} exploit supervised learning to determine the \acrshort{ris} beamformer and the position of user.

The positioning systems and methods discussed above, albeit promising, still have deficiencies in their hardware and software designs, preventing them from being deployed to diverse practical environments.
\emph{Firstly}, in the hardware domain, most existing works have considered positioning using signals with rather limited bandwidth, resulting in deficient range resolution and low adaptivity to the frequency selectivity of diverse environments caused by multipath fading.
The limited operating bandwidth of existing designs is partially \revise{attributed to the physical implementation of \acrshort{me}s, which is intrinsically highly frequency selective~\cite{Boyarsky2021Electronically, Deng2021Reconfigurable}, leading to a severe beam-squinting problem for signals with wide and ultra-wide bandwidth~\cite{Wei2021Channel}.
This means that a single configuration of \acrshort{me}s cannot provide the desired beam patterns over a large bandwidth simultaneously as the \acrshort{me}s' signal radiation coefficients vary largely across the band in terms of their phases and amplitudes.
In this regard, multi-band~(\acrshort{mb}) transmission~\cite{Elbahhar2012Indoor,Noschese2021Multi} is a promising alternative to ultra-wideband transmission.
\frev{The feasibility of MB-RHSs has been verified in~\cite{Boyarsky2021Electronically}, where an MB-RHS capable of operating in bands at $9.5$, $10$, $10.5$, and $11$~GHz is realized.
Moreover, in~\cite{Lin2022JPD_Dual,Zhang2020JESTCS_Programmable}, the authors prototyped RISs employing meta-elements capable of operating in two bands.}}
Exploiting \acrshort{mb} transmissions, \acrshort{rhs}-based \acrshort{bs}s can leverage a larger bandwidth while concurrently realizing appropriate beam patterns.

\emph{Secondly}, in the software domain, most positioning methods rely on raw received signals or extracted features such as time-of-arrival~(ToA) and angle-of-arrival~(AoA). 
Such methods lack environmental adaptivity as they cannot fully exploit the environment-specific features contained in the received signals for precision maximization~\cite{huang2020holographic}.
Although a few recent proposals have started to leverage deep learning techniques for automatic and environmental-specific feature selection and extraction~\cite{nguyen2020reconfigurable,Hu2021MetaSensing}, they need massive data on received signals and position labels for training.
To obtain the required data and labels, \emph{crowd sensing techniques}, where a crowd of users gather the data collaboratively~\cite{elzanaty2021towards, Han2016Truthful}, can be potentially exploited.
Nevertheless, it is nontrivial to provide effective incentives for users to disclose their position labels since these labels can indicate personal interests and hence potentially compromise user privacy~\cite{He2015User,Decker2008Location}.

\begin{figure}[t]
\centering
\includegraphics[width=1\linewidth]{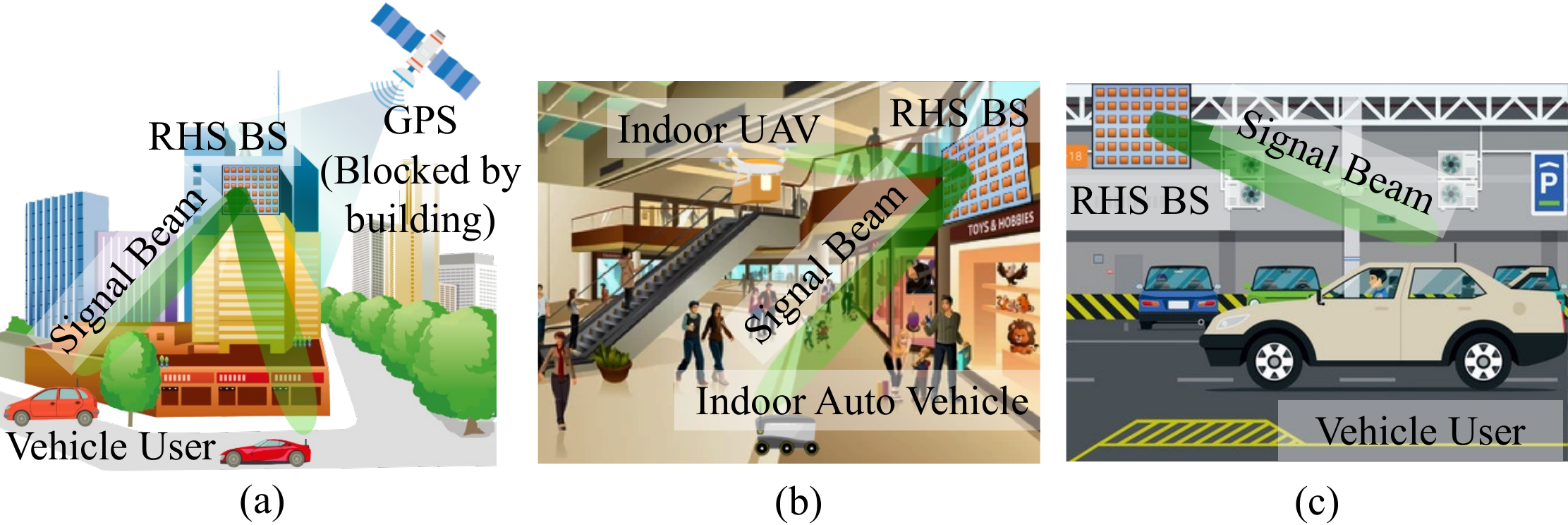}
\vspace{-1.5em}
\caption{Application scenarios of \gls{SysName}. (a) Outdoor vehicle positioning for GPS-deprived regions. (b) \revise{Indoor autonomous vehicle (e.g., \acrlong{uav}s~(\acrshort{uav}s) and cleaning robots) positioning for large malls.} (c) Underground vehicle positioning for garage parking.}
\label{fig: intro fig-1}
\vspace{-.5em}
\end{figure}

To achieve high position precision and environmental adaptivity while preserving user privacy, we propose \emph{\gls{SysName}}, an \acrshort{isac} system with positioning capability specifically targeting a wide range of outdoor, indoor, and underground GPS-deprived scenarios, as shown in Figs.~\ref{fig: intro fig-1}(a)-(c).
\gls{SysName} exploits \acrshort{mb}-\acrshort{rhs} and \acrfull{fl} to provide the environmental adaptivity in the hardware and software domains, respectively.
\revise{We propose a positioning protocol under the \acrshort{fl} framework for \gls{SysName}, allowing the users to collaboratively adapt \gls{SysName} to diverse environments while preserving the privacy of their own position labels.
With the proposed protocol, the \acrshort{bs} first transmits signals to the users, utilizing the \acrfull{dna} beamforming capability of \acrshort{mb}-\acrshort{rhs}.
Each user then employs a \emph{position estimator} function distributed by the \acrshort{bs} to process the received signals and estimate its position.
Furthermore, each user trains the position estimator with its local data, and the \acrshort{bs} schedules the users to send the trained position estimator in the uplink to perform global updates.}

To optimize the performance of \gls{SysName}, we derive a lower bound on the mean squared error~(\acrshort{mse}) of positioning considering the influence of \acrshort{mb} multipath fading; this bound is then exploited to optimize the \acrshort{dna} beamforming.
Besides, to facilitate \acrshort{fl} in practice where users have few position labels, we exploit the transfer learning technique to handle the insufficient training data. 
Moreover, a user scheduling algorithm is designed for \acrshort{fl}, jointly considering the convergence and efficiency.
The main contributions of this paper can be summarized as follows:
\begin{itemize}[leftmargin=12pt]
\item We propose the first positioning system assisted by both \acrshort{mb}-\acrshort{rhs} and \acrshort{fl}, delivering low positioning error and high environmental adaptivity without compromising the users' privacy.
\item \revise{We derive a lower bound on \gls{SysName}'s positioning error variance and utilize it for the optimization of \acrshort{dna} beamforming of \acrshort{mb}-\acrshort{rhs}s. Besides, in \acrshort{fl}, we handle the lack of users' position labels by exploiting the transfer learning technique and design a user scheduling algorithm to optimize the convergence and efficiency of the training.}
\item We verify the effectiveness of \gls{SysName} through
extensive simulations. 
\revise{Our results confirm that the proposed algorithm is more efficient compared to two benchmarks for beamforming optimization and user scheduling.}
The results also demonstrate that \gls{SysName} can effectively adapt to diverse environments and achieve low positioning errors.
\end{itemize}

Compared to its conference version~\cite{Hu2023ICC}, this paper proposes to apply FL for achieving privacy-preserving environmental adaptivity.
Furthermore, it provides new optimization algorithms, which enhance the efficiency of the DA beamforming optimization by proximal stochastic descent, handle the insufficiency of users' local data by transfer learning, and improve the efficiency of user scheduling in FL based on an new analytical result of the convergency rate.
The remainder of this paper is organized as follows:
In Sec.~\ref{sec: system model}, the system model for \gls{SysName} is established.
Then, for positioning error minimization, an optimization problem is formulated in Sec.~\ref{sec: prob formulate and decompose}.
In Sec.~\ref{sec: alg design}, we propose an efficient algorithm for \gls{SysName} to solve the formulated problem.
In Sec.~\ref{sec: evaluation}, simulation results are provided, and conclusions are drawn in Sec.~\ref{sec: conclu}.

\emph{Notations}: $\overline{(\cdot)}$, $(\cdot)^{\top}$, $(\cdot)^{\hil}$, and $(\cdot)^{-1}$ are the conjugate, transpose, Hermitian  transpose, and inverse operators, respectively. 
$\odot$ and $\otimes$ denote the Hadamard and Kronecker products, respectively. 
$\mathbb R^{M\times N}$ and $\mathbb C^{M\times N}$ denote the sets of real and complex $M\times N$ matrices, respectively. 
$\bm 1_{M}$ is the $M$-dimensional all-ones column vector, and $\bm 1_{M\!\times\! N}$ is the $M\times N$ all-ones matrix.
Functions $\trace(\cdot)$ and $\diag(\cdot)$ return the trace and the main diagonal vector of a matrix, respectively.
Function $\mathbb E_{\bm x\sim \varGamma}(\cdot)$ returns the expectation of the argument, given variable $\bm x$ follows distribution $\varGamma$.
Operators $\|\cdot \|_1$ and $\|\cdot\|_2$ are the $\ell_1$- and $\ell_2$-norms, respectively.
$\nabla_{\bm x} f$ represents the gradient vector of function $f$ with respect to $\bm x$.
Symbol $\iu$ is the imaginary unit.
$[\bm x]_{m}$, $[\bm X]_m$, and $[\bm X]_{m,n}$ are the $m$-th element of vector $\bm x$, the $m$-th row vector of matrix $\bm X$, and the $(m,n)$-th element of matrix $\bm X$, respectively.
$\{\bm x_i\}_i$ is the set of $\bm x_i$ for all subscript $i$ within its range.
$\bm x^{\circ 2}$ is the element-wise square of $\bm x$.
$\Re(\cdot)$ is the real part of the argument.

\section[System Model]{System Model}
\label{sec: system model}

\begin{figure}[t]
\centering
\includegraphics[width=1\linewidth]{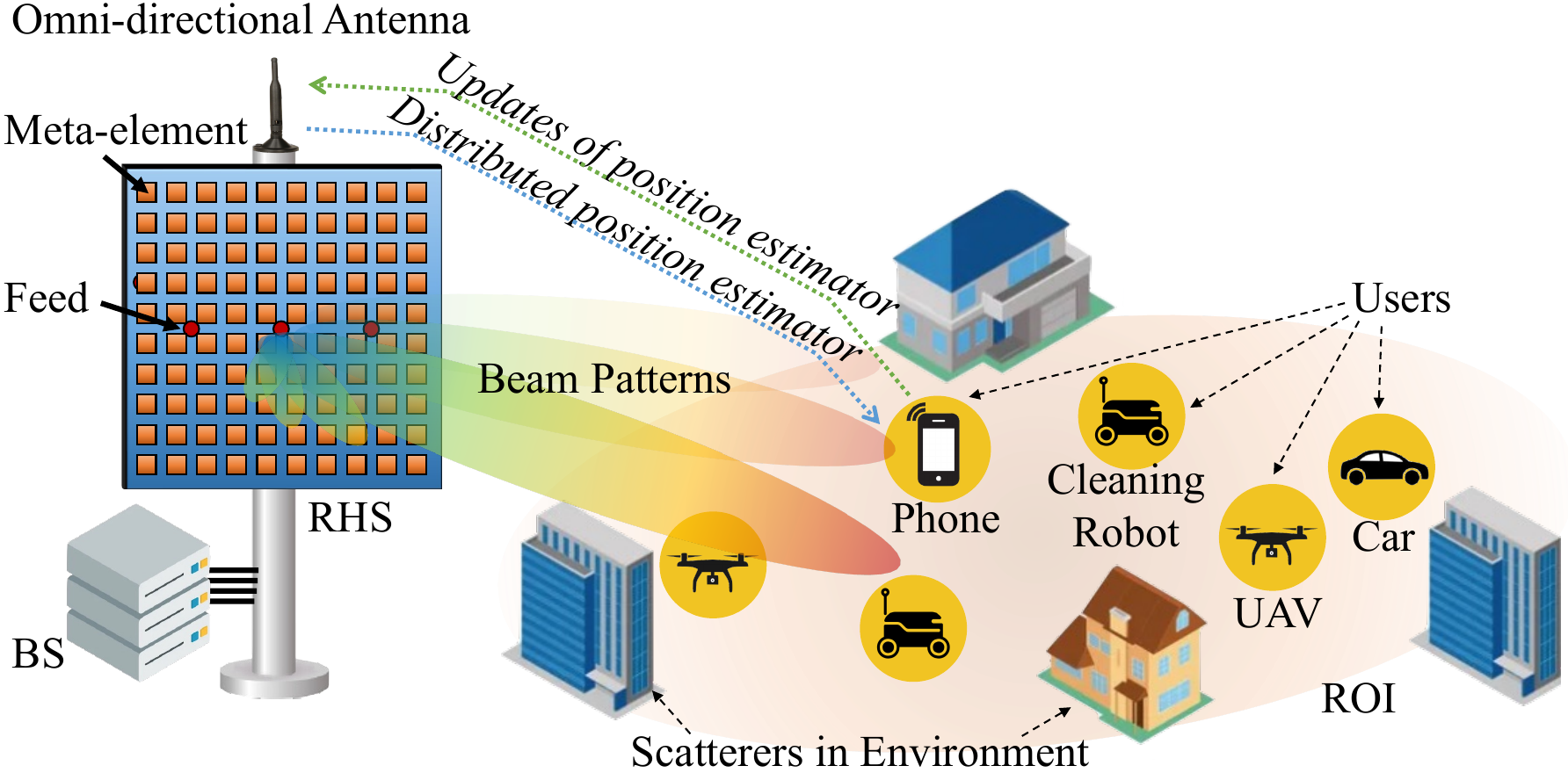}
  \caption{Illustration of the proposed \gls{SysName} system.}
  \label{fig: sys_mod fig-1}
\end{figure}

The proposed \gls{SysName} is an \acrshort{isac} system with positioning functionality exploiting an \acrshort{mb}-\acrshort{rhs} and \acrshort{fl}.
\revises{As in~\cite{Elzanaty2021Reconfigurable}, we assume the system utilizes the \acrfull{ofdm} waveform, which is typically adopted for \acrshort{isac} systems due to its high spectral efficiency, robustness against multipath fading, and easy implementation~\cite{Zhou2022Integrated}.}
As shown in Fig.~\ref{fig: sys_mod fig-1}, the system comprises a \acrshort{bs} equipped with an \acrshort{rhs} and $\gls{numUser}$ users.
\revise{Possible users include cars, autonomous vehicles~(UAVs and cleaning robots), and mobile phones.}
\revise{The \acrshort{bs} provides data and positioning services for the users in a time-division duplex~(TDD) manner.}
In this paper, we focus on developing the positioning function of \gls{SysName} for a 3D \acrfull{roi}.

Specifically, in \gls{SysName}, the process for users to obtain their positions is referred to as the \emph{positioning process}.
As the users' positions constitute private information, they are not intended to be known by the BS without users' explicit acknowledgement.
\revise{Thus, the users in \gls{SysName} estimate their positions by themselves, instead of relying on the \acrshort{bs} to estimate their positions and then inform them.}
\revise{This self-positioning is done by using a function referred to as \emph{\acrshort{pos_est}}.
Nevertheless, since it is hard for individual users to determine the exact characteristics of the \acrshort{rhs} and the environment, they cannot effectively derive the \acrshort{pos_est}.
To handle this issue, the \acrshort{pos_est} is provided to the users by the \acrshort{bs}.
Moreover, to make \gls{SysName} environment-adaptive, the users train the \acrshort{pos_est} collaboratively. 
For the positioning process, a \emph{federated positioning protocol} is proposed as detailed in this section.}

In the following, the components of the system are described in Sec.~\ref{ssec:syscomp}, the users' received signals are modeled in Sec.~\ref{s2ec: Received Signal Model}, and the federated positioning protocol is provided in Sec.~\ref{sec: pos protocol}.

\subsection{System Components} 
\label{ssec:syscomp}
The complete \gls{SysName} system comprises an \acrshort{rhs}-based \acrshort{bs} and multiple users.
\subsubsection[BS]{\acrshort{rhs}-based \acrshort{bs}}
\label{s3ec: sys mod bs}
The \acrshort{bs} is equipped with an \acrshort{rhs}, which is a rectangular planar antenna array composed of $\gls{numElem}$ reconfigurable \acrshort{me}s and $\gls{numFeed}$ signal feeds. 
Each \acrshort{me} can be modelled as an electronic controllable micro-antenna element, which takes the input \acrshort{rf} signals from the feeds and radiates the signals into space.
Besides, by electronically configuring the different \emph{state}s of the \acrshort{me}, its signal \emph{radiation coefficient} can be controlled, which determines the ratio between the signals emitted by and fed into it.
Based on~\cite{Zhang2022Holographic,Smith2017Analysis}, the radiation coefficient is assumed to be a real number from set\footnote{If \gls{setRadCoeff} is a discrete set, the method proposed in this paper can still be employed by adding an extra quantization step. } $\gls{setRadCoeff}= [0,1]$.

For positioning, the \acrshort{bs} transmits \acrshort{ofdm} signals in each of \gls{numBand} bands, where the spectral interval between different bands is assumed to be large to exploit spectrum diversity. 
Each band has \gls{numSBand} sub-bands, and the bandwidth of each sub-band is \gls{subBandwidth}.  
We assume that the radiation coefficient of a \acrshort{me} in a certain state is constant within the sub-bands of the same band as in~\cite{Zhang2022Holographic}.  
Nevertheless, due to the large spectral interval between different bands and the frequency selectivity of the \acrshort{me}s, the radiation coefficient of a \acrshort{me} in a certain state varies across different bands~\cite{Boyarsky2021Electronically, Deng2021Reconfigurable}.

\revise{Besides, as for communication, since we focus on the positioning function of \gls{SysName}, we assume a simple setting where the \acrshort{bs} employs an omnidirectional antenna and communicates with the users by single-band \acrshort{ofdm}}\footnote{\revise{While the \acrshort{rhs} can also be used for communication, this is beyond the scope of this paper. Furthermore, involving it will incur additional complexity for the design of \gls{SysName}'s positioning function, and thus it is not considered.}}.
\revise{With this antenna, the \acrshort{bs} broadcasts beaconing frames, receives data from the users, and sends control signals to the users.}

\subsubsection[User]{Users}
\label{s3ec: sys component user}
\revises{The users are assumed to have omnidirectional single Tx and Rx antennas, and they can communicate with the \acrshort{bs} in each of the \gls{numBand} bands. 
Nevertheless, due to bandwidth limitation, a user can only transmit or receive signals over one of the \gls{numBand} bands at a time.}
Besides, assuming each positioning process has a very short time duration, then the position of a user during a positioning process can be considered fixed and denoted by vector $\gls{userPos}\in \gls{tarSpace} \subseteq \mathbb R^{3}$ with \gls{tarSpace} denoting the \acrshort{roi}.
Moreover, we assume that the positioning processes take place periodically.  
In each positioning process, user positions are assumed to be independent and identically distributed random variables, each following a distribution $\gls{userDistribu}$, i.e., $\gls{userPos}\sim \gls{userDistribu}$.

\subsection{Received Signal Model}
\label{s2ec: Received Signal Model}

We establish the model for the received signals for the frame transmission from the  \acrshort{bs} to user $n$~($n\in\gls{valueSet_n}$), which is also referred to as the \emph{channel model}.
Without loss of generality, we index the frame by $q$, and omit subscript $n$ for the conciseness of presentation.
The signals transmitted by the \acrshort{bs} and received by the user undergo three stages of propagation, which are referred to as \emph{feed$\rightarrow$\acrshort{me}}, \emph{\acrshort{me} radiation}, and \emph{\acrshort{me}$\rightarrow$user}.

\subsubsection[Feed2Element]{Feed$\rightarrow$\acrshort{me}}
Denote the positions of feed~$k$ and \acrshort{me} $m$ by $\gls{posFeed_k}$ and $\gls{posElem_m}$~($\forall  k\in\gls{valueSet_k}, m\in \gls{valueSet_m}$), respectively.
Then, based on~\cite{Zhang2022Holographic} and~\cite{Elzanaty2021Reconfigurable}, for sub-band $j$ of band $i$ and frame $q$~($\forall i \in \gls{valueSet_i}, j\in\gls{valueSet_j}, q\in\gls{valueSet_q}$, \gls{numFrame} is the number of frames), the incident signals of \acrshort{me}~$m$ can be expressed as
\beq
\label{equ: gamma_ij}
\omega_{i,j,m}^{(q)} \!=\! \sum_{k=1}^{\gls{numFeed}} s_{i, j, k}^{(q)} \gls{direPatternFeed}(\bm\varphi^{\aod}_{k,m})   \kappa(f_{i,j}, \bm p^{\mathrm{F}}_{k}, \bm p^{\mathrm{E}}_{m})    \gls{direPatternElem}(\bm\varphi^{\aoa}_{k,m}),
\eeq
where $s_{i,j,k}^{(q)}$ denotes the \emph{digital symbol} transmitted by the \acrshort{bs} to the \gls{rhs} via feed $k$,
$ \gls{direPatternFeed}(\cdot)$ and $\gls{direPatternElem}(\cdot)$ represent the \emph{gain patterns} of the feed and the \acrshort{me}, respectively, 
$\bm\varphi^{\aod}_{k,m},\bm\varphi^{\aoa}_{k,m}\in\mathbb R^{2\times 1}$ are the \acrfull{aod} and the \acrfull{aoa} between feed~$k$ and \acrshort{me}~$m$, respectively, 
$f_{i,j}$ is the center frequency of sub-band~$j$ of band~$i$, 
and $\kappa(f_{i,j}, \bm p^{\mathrm{F}}_{k}, \bm p^{\mathrm{E}}_{m})$ represents the gain of the on-board propagation from $ \bm p^{\mathrm{F}}_{k}$ to $\bm p^{\mathrm{E}}_{m}$ at frequency $f_{i,j}$~(see~\cite{Zhang2022Holographic}):
\beq
\kappa(f_{i,j},  \bm p^{\mathrm{F}}_{k}, \bm p^{\mathrm{E}}_{m}) = \exp\Big( -\iu \cdot\frac{2\pi n_{\mathrm{r}}f_{i,j}}{v_0} \cdot \|\bm p^{\mathrm{E}}_{m} - \bm p^{\mathrm{F}}_{k}  \|_2 \Big).
\eeq
Here, $v_0$ is the speed of light and $n_{\mathrm{r}}$ is the refractive index of the \acrshort{rhs} board.
Moreover, in each frame $q$, the $\gls{numFeed}$ digital symbols transmitted by the \acrshort{bs} in each sub-band $j$ are bounded by a power constraint, i.e., $\sum_{k=1}^{\gls{numFeed}}\|s_{i,j,k}^{(q)}\|_2^2= P_{\max}$, where $P_{\max}$ is the maximum transmit power\footnote{\revise{Assuming $P_{\max}$ is fully utilized maximizes the received \acrshort{snr} of the users, which helps to minimize the positioning errors.}}.

\subsubsection[ElementRadiation]{\ecapitalisewords{\glsentryshort{me}} radiation}
\revise{Then, for frame $q$ and band $i$, the incident signals to each \acrshort{me}~$m$ are influenced by its radiation coefficient denoted by $\gls{code_iqm}$, which is assumed to be constant for the sub-bands of band $i$ as described in Sec.~\ref{s3ec: sys mod bs}.}
\revise{Thus, in sub-band $j$ of band $i$}, the radiated signals of \acrshort{me}~$m$ \revise{in frame $q$} can be expressed as
\begin{align}
\label{equ: holoVec_ijqm}
& \gls{holoVec_ijqm} = \gls{code_iqm} \cdot \omega_{i, j, m}^{(q)}.
\end{align}

\subsubsection[Element2User]{\ecapitalisewords{\glsentryshort{me}}$\rightarrow$user}
\label{s3ec: element 2 user channel}
The radiated signals are then received by the users.
\revise{For sub-band $j$ of band $i$}, the received signal of the user \revise{in frame~$q$} can be expressed as
\beq
\label{equ: received signal i j}
\gls{recvSig_ijq} = \sum_{m=1}^{\gls{numElem}}(\gls{losgain_ijm} + \gls{mpgain_ijmq}) \cdot \gls{holoVec_ijqm} + \gls{noise_ijq},
\eeq
where $\gls{noise_ijq}\sim \mathcal{CN}(0,\gls{noisePower})$ is the thermal noise following the complex Gaussian distribution with variance $\gls{noisePower}$, and $\gls{losgain_ijm}$ and $\gls{mpgain_ijmq}$ are the \acrfull{los} and multipath gains, respectively.
Denoting the power spectral density of the noise by \gls{noisePowerDensity}, the variance can be expressed as $\gls{noisePower} = \gls{noisePowerDensity}\gls{subBandwidth}$.
We note that, in~(\ref{equ: received signal i j}), the \acrshort{bs} and the users are assumed to be fully synchronized as in~\cite{Elzanaty2021Reconfigurable}.
Then, based on the signal propagation model in~\cite{Dardari2022LOS} and~\cite{Tang2021Wireless}, $\gls{losgain_ijm}$ can be modelled as
\beq
\gls{losgain_ijm}= \frac{v_0 \cdot \gls{direPatternElem}(\bm\theta_m^{\aod}) \cdot \gls{direPatternUser} }{4\pi f_{i,j} \cdot \|\gls{userPos} - \gls{posElem_m}\|_2} \cdot \exp\! \left(
-\iu \frac{2\pi f_{i,j}}{v_0} \|\gls{userPos} - \gls{posElem_m}\|_2
\right).
\eeq
Here, \gls{direPatternUser} denotes the gain of the user's Rx antenna for sub-band $j$ of band $i$. 
The Rx antennas of the $\gls{numUser}$ users are assumed to have the identical gains.
Besides, $\bm\theta_m^{\aod}\in\mathbb R^{2\times 1}$ is the \acrshort{aod} of the signals from \acrshort{me}~$m$ to the user.

Based on~\cite{Yu2004Modeling, goldsmith2005wireless}, we model multipath gains as complex Gaussian random variables satisfying \acrlong{wss} condition.
\revise{Defining $ \gls{mpgainVec_ijq}= (h^{\mulpath,(q)}_{i,j,1}, \cpdots, h^{\mulpath,(q)}_{i,j,\gls{numElem}})^\tran$, based on~\cite{Barriac2006Space}, $\gls{mpgainVec_ijq}\sim \mathcal{CN}(\bm 0, \gls{covmat_i})$}, where covariance matrix $\gls{covmat_i}\in\mathbb C^{\gls{numElem}\times\gls{numElem}}$ can be derived from the expectation of the outer product of the \acrshort{rhs}'s array response $\gls{angularResponse_i}(\bm \theta)\in \mathbb C^{\gls{numElem}}$ over the angular domain\footnote{
\red{Here, we assume that $\gls{covmat_i}$ only depends on the multi-band index $i$ since the multipath gains satisfy the \acrlong{wss} condition, and the sub-band frequencies are close to the center frequency of band $i$.}}, i.e., 
\beq
\label{equ: cov mat of multipath}
\begin{split}
\bm V_i = \mathbb E \left( 
\gls{angularResponse_i}(\bm \theta) \gls{angularResponse_i}(\bm \theta)^{\hil} \right) 
=\!\oint \gls{angularResponse_i}(\bm \theta) \gls{angularResponse_i}(\bm \theta)^{\hil}\! P_{\mathrm{pap},i}(\bm \theta) \dd \! \bm \theta.
\end{split}
\eeq
Here, $[\gls{angularResponse_i}(\bm \theta)]_{m} = \exp(\iu \frac{2\pi f_i}{v_0} (\gls{posElem_m} - \gls{posElem_1}) \cdot \hat{\bm n}(\bm \theta)) \cdot g^{\elem}_i(\bm \theta)$, where $f_i$ is the center frequency of band~$i$, $\hat{\bm n}(\bm \theta)$ is the unit normal vector for $\bm\theta$, and $g^{\elem}_i(\cdot)$ is the gain pattern of a \acrshort{me} at $f_i$.
Besides, $P_{\mathrm{pap},i}(\bm \theta)$ is the \emph{power-angle profile}~\cite{goldsmith2005wireless}, which accounts for the angular distribution of multipath gains.
\srev{We note that $\bm V_i$ can also account for the passive interference among users, i.e., the interference caused to a given user by signals passively scattered by the bodies of other users; because the scattering paths can be modelled as random multipath components.}

Based on~\cite{Barriac2006Space}, we model the covariance matrix between the multipath gain vectors for different frames and sub-bands of band $i$ as
\begin{align}
&\mathbb E\!\left(\bm h_{i, j_1}^{\mulpath, (q_1)} \big(\bm h_{i, j_2}^{\mulpath, (q_2)}\big)^{\hil}\right)\! =\! \gls{covCoefFuncFreq_i}(j_1, j_2)\cdot  \gls{covCoefFuncTime_i}(q_1, q_2)\cdot \bm V_i, \nonumber \\
&\quad \forall j_1,j_2\in\gls{valueSet_j} \text{ and }q_1, q_2 \in\gls{valueSet_q}, 
\end{align}
where $\gls{covCoefFuncFreq_i}(j_1, j_2)$ and $\gls{covCoefFuncTime_i}(q_1, q_2)$ denote the \emph{coherence coefficients} of sub-bands $j_1$ and $j_2$ and frames $q_1$ and $q_2$, respectively.
Based on~\cite{Barriac2006Space}, they can be expressed as follows:
\beq
\label{equ: covCoefFuncFreq}
\begin{split}
& \gls{covCoefFuncFreq_i}(j_1, j_2) = \frac{1}{1+\iu 2\pi \gls{rmsValue_i} (f_{i,j_1} - f_{i,j_2})}, \\
& \gls{covCoefFuncTime_i}(q_1, q_2) = J_0(2\pi \gls{dopplerFreq_i}\revise{\gls{durFrame}\cdot(q_1 - q_2)}),
\end{split}
\eeq
\revise{where \gls{durFrame} denotes the duration of a frame}, $\gls{rmsValue_i}$ denotes the \emph{\acrfull{rms} power delay spread} of band~$i$, $J_0(\cdot)$ is the \emph{zeroth-order Bessel function of the first kind}, and $\gls{dopplerFreq_i} = \gls{userMaxSpeed} f_i / v_0$ is the maximum Doppler frequency with $\gls{userMaxSpeed}$ being the users' maximum speed.

\red{Moreover, as the spectral intervals between different \acrshort{ofdm} bands are large, the multipath gain vectors of different bands are assumed to be not correlated}, i.e.,
\beq
\label{equ: zero correlated}
\mathbb E\left(\bm h_{i_1, j_1}^{\mulpath, (q_1)}\big(\bm h_{i_2, j_2}^{\mulpath, (q_2)}\big)^{\hil}\right) = \bm 0,~\forall i_1\neq i_2.
\eeq

In summary, for the $\gls{numFrame}$ transmitted frames, we can arrange the digital symbols in~(\ref{equ: gamma_ij}) transmitted in sub-band $j$ of band $i$ in a matrix $\bm S_{i,j}\in\mathbb C^{\gls{numFrame}\times\gls{numFeed}}$ with $[\bm S_{i,j}]_{q,k} = s_{i,j,k}^{(q)}$ and arrange the radiation coefficients in~(\ref{equ: holoVec_ijqm}) for band $i$ in a matrix $\bm C_i\in\mathbb C^{\gls{numFrame}\times \gls{numElem}}$ with $[\bm C_i]_{q,m} = c_{i,m}^{(q)}$.
Since $\{\bm S_{i,j}\}_j$ and $\bm C_i$ control the \acrshort{dna} beamforming, we refer to them as the \emph{\acrshort{dna} beamforming configuration for band $i$}.
Based on~(\ref{equ: gamma_ij}),~(\ref{equ: holoVec_ijqm}), and~(\ref{equ: received signal i j}), the received signals of a user for band $i$ are collected in vector $\bm y_{i}(\cdot)\in\mathbb R^{\gls{numFrame}\gls{numSBand}\times1}$, which is a function of $\bm p$, $\{\bm S_{i,j}\}_j$, and $\bm C_i$, and can be expressed as
\beq
\label{equ: recvSigVec_i}
\bm y_{i}(\gls{userPos}; \{\bm S_{i,j}\}_j, \bm C_{i}) =  \diag\big(\big(\bm  H_i^{\los} \otimes \bm 1_{\gls{numFrame}} + \bm H_i^{\mulpath}\big) \gls{holoMat_i}^\tran \big)  + \bm e_i.
\eeq
Here, $\bm e_i\sim\mathcal{CN}(0, \gls{noisePower}\bm I_{\gls{numFrame}\gls{numSBand}})$ is the noise vector, and the elements of the matrices appearing in~(\ref{equ: recvSigVec_i}) can be expressed as follows~($\forall j\!\in\! \gls{valueSet_j}, k \!\in\! \gls{valueSet_k}, m \!\in\! \gls{valueSet_m}, q\!\in\!\gls{valueSet_q}$):
\begin{align}
\label{equ: matrix forms elements}
& [\bm H_i^{\los}]_{j, m} \!=\! \gls{losgain_ijm},~[\bm H_i^{\mulpath}]_{(q-1)\gls{numSBand}\!+\!j, m} \!=\! \gls{mpgain_ijmq}, \nonumber\\
&[\bm B_{i,j}]_{k,m} =\gls{direPatternFeed}(\bm\varphi^{\aod}_{k,m})\cdot \gls{direPatternElem}(\bm\varphi^{\aoa}_{k,m})  \cdot \kappa(f_{i,j}, \bm p^{\mathrm{F}}_{k}, \bm p^{\mathrm{E}}_{m}),\nonumber\\
&[\bm T_i]_{j} = \bm C_i \odot \big( \bm S_{i,j} \bm B_{i,j} \big),
\end{align}

\textbf{Remark 1}: 
In the established model, several parameters are highly sensitive to the hardware implementation and environment and hard to obtain precisely.
For instance, the actual gain pattern of the \acrshort{me}, i.e., $\gls{direPatternElem}(\bm\theta_m^{\aod})$, generally differs from the theoretical model due to unexpected imperfections in the implementation.
Besides, the power-angle profile, i.e., $P_{\mathrm{pap},i}(\bm \theta)$, is hard to obtain due to the complex influence of signal scatterers in diverse \acrshort{roi}s.
We refer to these parameters as the \emph{environmental characteristics}.
\gls{SysName} achieves adaptivity to the environmental characteristics via a federated positioning protocol, which is introduced next.

\subsection[Federated Positioning Protocol]{Federated Positioning Protocol}
\label{sec: pos protocol}

\begin{figure*}[t]
\centering
\includegraphics[width=.8\linewidth]{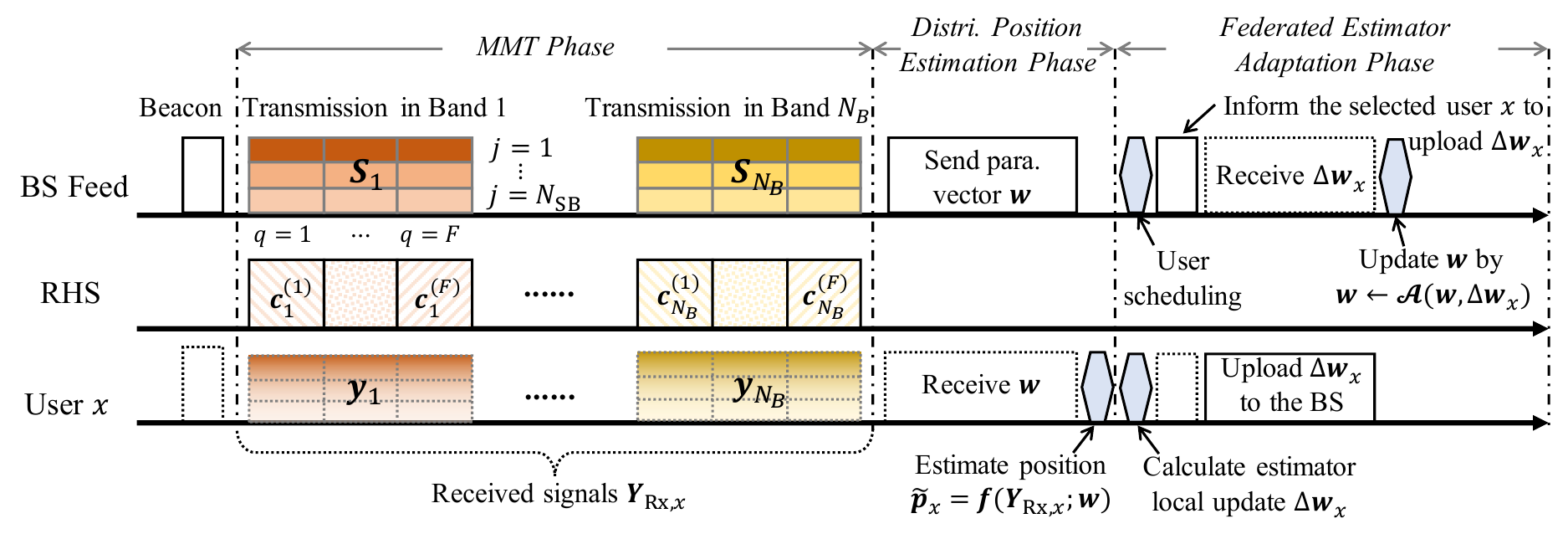}
\vspace{-.5em}
\caption{Federated positioning protocol for \gls{SysName}. \revise{The index of the positioning process, $t$, is omitted to facilitate presentation.}}
\vspace{-.5em}
\label{fig: sys_mod fig-2}
\end{figure*}

\revise{When a user needs to estimate its position, it requests the \acrshort{bs} to conduct a positioning process.
After receiving the request, the \acrshort{bs} broadcasts a beaconing frame informing the users of the beginning of the positioning process.
To coordinate the positioning process and enable \gls{SysName} to adapt to diverse environments, we propose a federated positioning protocol.}
As shown in Fig.~\ref{fig: sys_mod fig-2}, each positioning process is comprised of three phases, i.e., a \emph{\acrshort{mb} multi-pattern transmission~(MMT) phase}, a \emph{distributed position estimation phase}, and a \emph{federated adaptation phase}.
Without loss of generality, we index the positioning process by $t=0,\cpdots, \gls{numPP}$, and describe the $t$-th positioning process in the following.

\subsubsection[First Phase]{\acrshort{mmht} Phase}
\label{s3ec: MMHT phase}

\revise{In this phase, the \acrshort{bs} generates multiple beam patterns in each band by using the \acrshort{rhs}, providing users with distinct received signals for position estimation.
Since the \acrshort{bs} does not know the environmental characteristics or the users' positions, it does not distinguish between different positioning processes, and thus the generated beam patterns are independent of $t$.}
Besides, as described in Sec.~\ref{s3ec: sys mod bs}, the radiation coefficient of a \acrshort{me} in a certain state varies across different bands.
\frev{Thus, a state configuration that creates a desired beamforming pattern in one band may lead to undesired beam patterns in other bands.
Therefore, to design favorable beam patterns in all bands, we assume that the \acrshort{ofdm} transmissions in the \gls{numBand} bands are performed sequentially, allowing the states of the \acrshort{me}s to be configured independently in each band.
This approach also reduces the hardware requirements for the BS and the users as their \acrshort{rf} chains do not have to support ultra-wideband signal transmission and reception}\footnote{
\frev{If the RHS can independently control the beam patterns in multiple bands, \gls{SysName} can be modified to account for parallel transmissions in these bands, assuming the hardware of both the BS and users is capable of supporting it.}}.

\revise{In each band $i$, the \acrshort{bs} transmits \gls{numFrame} frames as shown in Fig.~\ref{fig: sys_mod fig-2}.}
\revise{Then, for positioning process $t$, the received signals of user~$n$ in all the \gls{numBand} bands are arranged in matrix $\gls{recvSigMat_n^t}\in \mathbb R^{\gls{numBand}\times\gls{numFrame}\gls{numSBand}}$, whose $i$-th row is $[\gls{recvSigMat_n^t}]_i = \bm y_i(\gls{userPos_n^t}; \{\bm S_{i,j}\}_j, \bm C_{i})$ based on~(\ref{equ: recvSigVec_i}), with $\gls{userPos_n^t}$ denoting user $n$'s position.}
\revise{To facilitate the presentation, we refer to $\{\bm S_{i,j}\}_{j}$ and $\bm C_{i}$ for all the \gls{numBand} bands collectively as the \emph{\acrshort{dna} beamforming configuration}, denoted by $\{\bm S_{i,j}\}_{i,j}$ and $\{\bm C_{i}\}_i$, which has an impact on \gls{SysName}'s positioning precision since it determines the beam patterns probing the \acrshort{roi}.}
\srev{Based on Fig.~\ref{fig: sys_mod fig-2}, the \acrshort{mmht} phase has linear time complexity with respect to~(w.r.t.) the number of bands and the number of frames transmitted in each band. 
Besides, since the BS broadcasts the frames to all users at the same time, the \acrshort{mmht} phase has constant time complexity w.r.t. the number of users.
Consequently, its time complexity is given by $\mathcal O(\gls{numBand}\gls{numFrame})$.}

\subsubsection[Second Phase]{\revise{Distributed Position 
Estimation Phase}}
\label{s3ec: protocol second phase}

In this phase, the \acrshort{bs} first distributes the \gls{pos_est} to the users through downlink beacon transmission.
\revise{The \gls{pos_est} is modeled as a multi-layer perceptron~(\acrshort{mlp}), which is a universal function approximator with high generalization capability~\cite{Chen1995Universal}.
The \gls{mlp} can be interpreted as a parameterized function with parameter vector $\bm w \in \mathbb R^{\gls{dimCoeff}}$, where $\bm w $ is comprised of the \gls{dimCoeff} connection weights and biases of the \gls{mlp}.
Specifically, in positioning process $t$, the distributed \acrshort{pos_est} can be denoted by $\bm f(\cdot ;\bm w^{(t)}):\gls{recvSigMat_n^t}\rightarrow \gls{userPos_est_n^t}$ with $\gls{userPos_est_n^t}\in\mathbb R^{3}$ denoting the estimated position of user $n$.}

\subsubsection[Third Phase]{\revise{Federated Adaptation Phase}}
\label{s3ec: fed ada phase}
Due to the unknown environmental characteristics, the \acrshort{bs} cannot effectively determine the \gls{pos_est} by itself.
Thus, in this phase, the users help the \acrshort{bs} to adapt the \gls{pos_est} to the environment by using their local datasets.
Here, the local dataset of a user contains the received signal matrices and their corresponding position labels.
To enable this adaptation while protecting the privacy of the users' position labels, the \emph{\acrshort{fl} framework} is employed, where the \gls{pos_est} is trained in a distributed manner with no position labels sent to the \acrshort{bs}.
Specifically, each user first calculates the gradient of the positioning error for its local dataset w.r.t. $\bm w^{(t)}$.
Then, the \acrshort{bs} schedules users to upload their gradients, and updates the \gls{pos_est} based on the received gradients.

More specifically, denote the local dataset gathered by user $n$ by $\mathcal D_n$, and assume that each user has obtained the data-label pairs in $\mathcal D_n$, i.e., $(\gls{recvSigMat}, \gls{userPos})\in \mathcal D_n$, when it was near a few \emph{anchors} in the \acrshort{roi}\footnote{%
\srev{Here, an anchor refers to a location where the users can obtain their position labels based on short-range positioning techniques.
Such short-range positioning techniques can be readily supported by near-field communication~(NFC) of the users with the existing Internet of Things~(IoT) infrastructure~\cite{Ozdenizci2015Sensors_NFC, Kortuem2010IntCom_Smart}.}}.
Then, based on the error measure commonly used for positioning systems, e.g.,~\cite{He2020Large}, we assume that user $n$ evaluates its positioning error by the \acrshort{mse} loss, i.e.,
\beq
\label{equ: def hat L n}
\hat{\mathcal L}_n(\bm w^{(t)}) = \sum_{(\gls{recvSigMat}, \gls{userPos})\in \mathcal D_{n}} \| \gls{userPos} - \bm f(\gls{recvSigMat}; \bm w^{(t)}) \|_2^2.
\eeq
Based on~(\ref{equ: def hat L n}), the gradient of user $n$'s local loss can be calculated as $\bm g_n^{(t)} = \nabla_{\bm w} \hat{\mathcal L}_n(\bm w^{(t)})$, which is referred to as its \emph{local gradient}.
\frev{Exploiting the local gradient of the positioning loss w.r.t. the position labels collected near a small number of anchors, \gls{SysName} adapts its position estimator to the actual deployment environment for achieving large-range positioning with high precision.}

Moreover, to prevent local gradients from compromising position privacy, the \emph{\acrfull{dp} training} mechanism is employed in \acrshort{fl}, i.e., noises are added to the local gradients.
For user $n$~($\forall n \in \gls{valueSet_n}$), the noise term added to $\bm g_n^{(t)}$ is denoted by $\bm \varsigma_n^{(t)}\in\mathbb R^{\gls{dimCoeff}}$ and follows Gaussian distribution $\mathcal N(\bm 0, \gls{dpSigma_n2}\bm I)$.
Based on~\cite{abadi2016deep}, variance $\gls{dpSigma_n2}$ can be calculated as
\beq
\label{equ: dp sigma}
\gls{dpSigma_n2} = \frac{L^2}{\gls{dpEpsilon_n2}}{2\log(1.25/\gls{dpDelta})},
\eeq
\revise{where $\gls{dpEpsilon_n}$ represents the \emph{privacy leakage bound} of user $n$ in terms of differential privacy, $L$ denotes the Lipschitz constant of the local gradient which can be enforced by having each user rescale its local gradient to $L$ in terms of $\ell_2$-norm, and $\gls{dpDelta}\ll 1$ is a small constant ensuring \gls{dpSigma_n2} to be finite by allowing a violation probability of the privacy leakage bound.}
Consequently, the \emph{local update} that user $n$ prepares to upload can be expressed as $\Delta \bm w_n^{(t)} = - \bm g_n^{(t)} + \bm \varsigma_n^{(t)}$.

Furthermore, as \gls{SysName} also needs to provide communication services, we assume that in each positioning process, only one user is selected to upload its local update over a single band%
\footnote{
\frev{The band used for uploading can be selected by the user for rate maximization. Even multiple bands can be used if the user and the BS can support it. The proposed algorithm can be modified to accommodate such cases, as described in Sec.~\ref{s2ec: alg design ftl}.}}, so that the occupation of the time-spectrum resources for \acrshort{fl} is minimized.
For the $t$-th positioning process, denote the probability of selecting each user for uploading by \emph{scheduling probability vector} $\bm \xi^{(t )} = (\xi^{(t)}_1, \cpdots, \xi^{(t)}_{\gls{numUser}})$. The update of the parameter vector can be expressed as
\beq
\bm w^{(t+1)} = \bm{\mathcal A}^{(t)}(\bm w^{(t)}, \Delta \bm w_x^{(t)})\big|_{x\sim \mathcal M(\bm \xi^{(t )})},
\eeq
where $x$ is the index of the selected user, $\bm {\mathcal A}^{(t)}(\cdot)$ denotes the \emph{adaptation function} used by the \acrshort{bs} to update $\bm w^{(t)}$ based on $\Delta \bm w_x^{(t)}$, and $\mathcal M(\bm \xi^{(t )})$ denotes the \emph{multinomial distribution} given $\bm \xi^{(t )}$.

\srev{\textbf{Remark 2}: In \gls{SysName}, users do not suffer from active user interference caused by signal transmissions of other users because they only receive signals in the \acrshort{mmht} phase and are scheduled to transmit their local updates one at a time in the federated adaptation phase.%
}

\section{Problem Formulation for Positioning Error Minimization}
\label{sec: prob formulate and decompose}

We formulate an optimization problem for \gls{SysName}, targeting the minimization of the average \acrshort{mse} of positioning experienced by the users over the \acrshort{roi} after adaptation.
The degrees of freedom for optimization include the \acrshort{dna} beamforming configuration, i.e., $\gls{sigSet}$ and $\gls{codeSet}$, the initial parameter vector of the \gls{pos_est}, $\bm w^{(0)}$, and the sets of adaptation functions and scheduling probability vectors for the $t = 0,\cpdots, T$ positioning processes, i.e., $\{\bm{\mathcal A}^{(t)}\}_{t}$ and $\{\bm \xi^{(t)}\}_{t}$.
The positioning error minimization problem is formulated as:
\begin{subequations}
\begin{align}
\label{opt P1: obj func}
\text{(P1):} \hspace{-.5em} & \min_{\substack{\gls{sigSet}, \gls{codeSet},\\ \bm w^{(0)}, \{\bm{\mathcal A}^{(t)}\}_{t}, \{\bm \xi^{(t)}\}_t}} 
 \sum_{n=1}^{\gls{numUser}} \mathop{\mathbb E}_{{\bm p}_n\sim \gls{userDistribu}}\hspace{-.3em}\left(\|{\bm p}_n - \bm f(\bm Y_{\mathrm{Rx}, n};\bm w^{(T)})\|_2^2\right)\!,\\
\label{opt P1: cons 1}
\text{s.t.} ~%
& [\bm Y_{\mathrm{Rx}, n}]_i = \bm y_i({\bm p}_n; \{\bm S_{i,j}\}_j, \bm C_{i}),\\
\label{opt P1: cons 2}
& \diag(\bm S_{i,j}\bm S_{i,j}^\hil) = P_{\max}\bm 1_{\gls{numFrame}},\\
\label{opt P1: cons 3}
& [\gls{codeMat_i}]_{q,m} \in \gls{setRadCoeff},\\
\label{opt P1: cons 4}
& \Delta \bm w_n^{(t)} = -\nabla_{\bm w} \hat{\mathcal L}_n(\bm w^{(t)}) + \bm \varsigma_n^{(t)},\\
\label{opt P1: cons 5}
& \bm w^{(t+1)} = \bm{\mathcal A}^{(t)}(\bm w^{(t)}, \Delta \bm w_x^{(t)})\big|_{x\sim \mathcal M(\bm \xi^{(t )})},\\
\label{opt P1: cons 6}
& \bm 0\preceq\bm \xi^{(t)}\preceq\bm 1, ~\|\bm \xi^{(t)}\|_1 = 1.\\
& \forall i\in\gls{valueSet_i},~j\in\gls{valueSet_j},~q\in\gls{valueSet_q},\nonumber\\[-.2em]
& ~m\in\gls{valueSet_m},~n\in\gls{valueSet_n},~t\in\gls{valueSet_t}, \nonumber 
\end{align}
\end{subequations}

\revise{In~(\ref{opt P1: obj func}), the objective function, denoted by $\mathcal L(\bm w^{(T)})$, is the expected \acrshort{mse} of positioning experienced by the $\gls{numUser}$ users after the adaptation in the $T$ positioning processes}.
\revise{Constraint~(\ref{opt P1: cons 1}) indicates that the relationship between the user position and the received signal matrix follows the channel model established in~(\ref{equ: recvSigVec_i}).
Constraints~(\ref{opt P1: cons 2}) and (\ref{opt P1: cons 3}) indicate that the symbols for digital beamforming satisfy the power constraint, and that the radiation coefficient value of each \acrshort{me} belongs to set~$\gls{setRadCoeff}$, respectively.
Besides, constraints~(\ref{opt P1: cons 4}) and~(\ref{opt P1: cons 5}) follow from the update of the parameter vector of the \acrshort{pos_est} in each positioning process according to the protocol proposed in Sec.~\ref{s3ec: fed ada phase}.
Moreover, constraint~(\ref{opt P1: cons 6}) ensures that $\bm \xi^{(t)}$ is a valid probability vector.}

\revise{The challenges of solving~(P1) comprise the following three aspects:
\emph{Firstly}, as described in the remark in Sec.~\ref{s2ec: Received Signal Model}, channel models $\{\bm y_i(\cdot)\}_i$ in~(\ref{opt P1: cons 1})  contain undetermined environmental characteristics, making them hard to evaluate the influence of the \acrshort{dna} beamforming configuration on the received signals. 
This hinders the optimization of \gls{sigSet} and \gls{codeSet}.}

\revise{\emph{Secondly}, in~(\ref{opt P1: cons 4}), as the local datasets only contain the position labels collected near a few anchors, the loss function of user $n$, $\hat{\mathcal L}_n(\cdot)$, is not identical to the objective function, $\mathcal L(\cdot)$. 
Thus, the update of the parameter vector in~(\ref{opt P1: cons 5}) does not necessarily reduce the expected \acrshort{mse} of positioning.
To handle this problem, initial parameter vector $\bm w^{(0)}$ and adaptation function set $\{\bm{\mathcal A}^{(t)}\}_t$ need to be properly selected.
However, the degrees of freedom of $\bm w^{(0)}$ and $\{\bm{\mathcal A}^{(t)}\}_t$ are very high due to the large number of parameters in the \acrshort{mlp} and the arbitrary forms of the adaptation functions, making the search for $\bm w^{(0)}$ and $\{\bm{\mathcal A}^{(t)}\}_t$ very challenging.}

\emph{Thirdly}, as the \acrshort{bs} does not know the users' local datasets or local gradients when determining the scheduling probability vectors, it cannot evaluate the influence of $\bm \xi^{(t)}$ let alone to optimize it.
Nevertheless, as $\{\bm \xi^{(t)}\}_t$ in~(\ref{opt P1: cons 5}) has an impact on the convergence and efficiency of \acrshort{fl}, it has a fundamental influence on the objective function.
Therefore, optimizing $\{\bm \xi^{(t)}\}_t$ is both crucial for solving (P1) and very challenging.

\revise{In summary, (P1) is highly complex and challenging, and beyond the capabilities of conventional optimization algorithms.
Hence, to solve~(P1), novel algorithms that can effectively handle the aforementioned three challenges are needed.}

\section{Positioning Error Minimization Algorithm for \gls{SysName}}
\label{sec: alg design}

\begin{figure*}[t]  
\centering
\includegraphics[width=.85\linewidth]{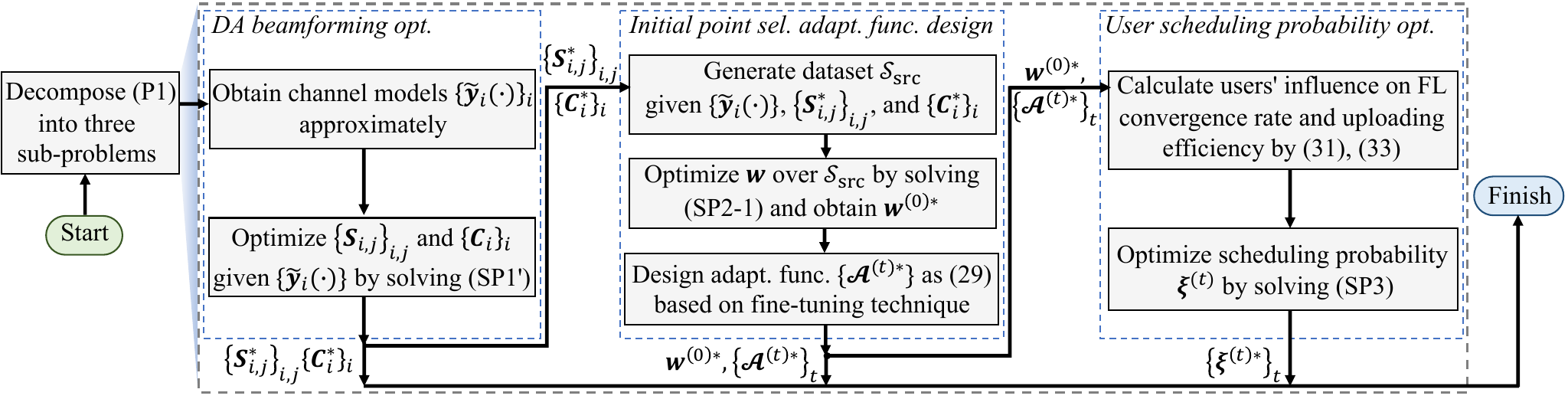}
\caption{\revise{Flow chart of the proposed positioning error minimization algorithm for \gls{SysName}.}}
\label{fig: alg flow chart}
\end{figure*}

\revise{In this section, we handle~(P1) by proposing a \emph{positioning error minimization algorithm for \gls{SysName}}.
Specifically, we tackle the three challenges arising when solving~(P1) by decomposing the problem into three sub-problems.
The three sub-problems are: the \emph{\acrshort{dna} beamforming optimization} by solving $\{\bm S_{i,j}^*\}_{i,j}$ and $\{\bm C_i^*\}_i$, the \emph{initial point selection and adaptation function design} for $\bm w^{(0)*}$ and $\{\bm{\mathcal A}^{(t)*}\}_{t}$, and the \emph{user scheduling probability optimization} by solving $\{\bm \xi^{(t)*}\}_t$.
A flow chart of the complete algorithm proposed for solving (P1) is provided in~Fig.~\ref{fig: alg flow chart}.}

\subsection{\acrshort{dna} Beamforming Optimization}
\label{s2ec: source domain solving alg}

\revise{To overcome the challenge due to undetermined environmental characteristics, we substitute the channel models in~(P1), $\{{\bm y}_i(\cdot)\}_i$, with deterministic ones denoted by $\{\tilde{\bm y}_i(\cdot)\}_i$, where the environmental characteristics such as the gain pattern of meta-elements and the power-angle profile are obtained approximately.
The reason why we can perform this substitution is that the federated positioning protocol enables \gls{SysName} to adapt to the environment, allowing the \acrshort{pos_est} to adjust to the difference between $\{{\bm y}_i(\cdot)\}_i$ and $\{\tilde{\bm y}_i(\cdot)\}_i$ via \acrshort{fl}.}
\revise{Thus, we can solve the \acrshort{dna} beamforming optimization problem given $\{\tilde{\bm y}_i(\cdot)\}_i$ and leave the adaptation to \acrshort{fl}.}
Besides, to facilitate the \acrshort{dna} beamforming optimization, we consider that the optimal \acrshort{pos_est} for \acrshort{dna} beamforming configuration is employed, which is denoted by $\tilde{\bm f}^*(\cdot)$.
Therefore, the optimization of \gls{sigSet} and \gls{codeSet} in (P1) is converted into a sub-problem:
\begin{align}
\text{(SP1):} &\min_{\gls{sigSet}, \gls{codeSet}}~
\mathop{\mathbb E}_{{\bm p}\sim \gls{userDistribu}}\! \left(\|{\bm p} - \tilde{\bm f}^*(\bm Y_{\mathrm{Rx}})\|_2^2\right)\!,\\
\text{s.t.} ~%
&[\bm Y_{\mathrm{Rx}}]_i = \tilde{\bm y}_i({\bm p}; \{\bm S_{i,j}\}_j, \bm C_{i}),~\text{(\ref{opt P1: cons 2})-(\ref{opt P1: cons 3})},~\forall i,j.\nonumber
\end{align}

Problem (SP1) is still challenging because: 
1) $\tilde{\bm f}^*(\cdot)$ is undetermined; 
2) the noises and multipath gains in the channel models are random variables;
3) the total dimension of \gls{sigSet} and \gls{codeSet} is very large, namely $\gls{numBand}\gls{numFrame}\gls{numSBand}\gls{numFeed}+\gls{numBand}\gls{numFrame}\gls{numElem}$, resulting in high computational complexity for evaluating the objective function of~(SP1) and its gradient.

To overcome the above challenges, we first convert~(SP1) into a \acrshort{crlb} minimization problem, addressing the challenges of the undetermined $\tilde{\bm f}^*(\cdot)$ and the random noises and multipath gains.
Then, to avoid the high computational complexity caused by the large dimension of \gls{sigSet} and \gls{codeSet}, the gradient of the \acrshort{crlb} is obtained in closed form, based on which an efficient stochastic gradient descent algorithm is proposed.
The detailed steps are described below.

\subsubsection{Deriving the \acrshort{crlb} on the \acrshort{mse} of Positioning}
\label{s3ec: deriving the crlb}
Given $\gls{sigSet}$ and $\gls{codeSet}$, we first analyze the objective function value of (SP1), which can be obtained by solving optimization problem
$\min_{\tilde{\bm f}(\cdot)}~\mathbb E_{(\gls{recvSigMat},\gls{userPos})} \big( \| \gls{userPos} - \tilde{\bm f}(\gls{recvSigMat})\|_2^2 \big)$.
Supposing the solution to the optimization problem, $\tilde{\bm f}^*(\cdot)$, is unbiased, the optimum of the objective in (SP1) can be benchmarked by the average \acrshort{crlb} over the \acrshort{roi}, since the \acrshort{crlb} is a valid lower bound for the variance of any unbiased estimator~\cite{kay1993fundamentals}.
Specifically, the \acrshort{crlb} for position~\gls{userPos} can be calculated based on the following Proposition~\ref{prop: crlb expression}.

\begin{proposition}
\label{prop: crlb expression}
	Given $\gls{sigSet}$ and $\gls{codeSet}$, the \acrshort{crlb} for a user at $\gls{userPos}$ can be calculated as
	\beq
\label{equ: general crlb expression}
\mathrm{CRLB}(\gls{userPos}) = \sum_{u=1}^3\big[
\bm I^{-1}_{\mathrm{FIM}} (\gls{userPos})
\big]_{uu}.
\eeq
Here, $\bm I_{\mathrm{FIM}}(\gls{userPos})$ is the \acrfull{fim} of $\gls{recvSigMat}$ w.r.t. $\gls{userPos}$, whose $(u,v)$-th element~($\forall u,\!v \!\in\! \{1,\!2,\!3\}$) is given by:
\begin{align}
\label{equ: fim matrix}
& [\bm I_{\mathrm{FIM}} (\gls{userPos})]_{u,v} \!=\!  2\Re\Big(\sum_{i=1}^{\gls{numBand}}\! \big(\frac{\partial \hat{\bm y}_i}{\partial p_u}\big)^{\hil}\! \bm \varLambda_i^{-1} \!\big(\frac{\partial \hat{\bm y}_i}{\partial p_v}\big) \Big), 
\end{align}
\revise{where $\hat{\bm y}_i$ denotes the expectation of $\tilde{\bm y}_i(\bm p)$, $p_u$ is the $u$-th element of $\gls{userPos}$, and $\bm \varLambda_i\in\mathbb C^{\gls{numFrame}\gls{numSBand}\times \gls{numFrame}\gls{numSBand}}$ is the covariance matrix of $\tilde{\bm y}_i$}. The terms in~(\ref{equ: fim matrix}) can be calculated as follows
\beq
\label{equ: some equ follow fim}
{\partial\hat{\bm y}_i \over \partial p_{u}} \!=\! \diag\big( (  \dot{\bm H}^{\los}_{i,u} \!\otimes\! \bm 1_{\gls{numFrame}} ) \gls{holoMat_i}^{\tran}\big),  \bm\varLambda_i \!=\! \bm K_{\mathrm{ft},i}\!\odot\! (\bm T_i \bm V_i \bm T_i^{\hil}) \!+\! \gls{noisePower}\bm I,
\eeq
where $\dot{\bm H}^{\los}_{i,u}= {\partial\bm H_i^{\los}/\partial p_u}$, and $\bm K_{\mathrm{ft},i}= ( \bm K_{\mathrm{f},i} \!\otimes\! \bm 1_{\gls{numFrame}\!\times\!\gls{numFrame}})\!\odot\! ( \bm 1_{\gls{numSBand}\!\times\!\gls{numSBand}}\!\otimes\!\bm K_{\mathrm{t},i} )$ with $[\bm K_{\mathrm{f},i}]_{j_1, j_2} =  \rho_{\mathrm{f},i}(j_1,  j_2)$ and $[\bm K_{\mathrm{t},i}]_{q_1, q_2} =  \rho_{\mathrm{t},i}(q_1, q_2)$.
\end{proposition}
\vspace{-.5em}
\begin{IEEEproof}
The \acrshort{fim} in~(\ref{equ: fim matrix}) can be obtained by substituting the channel model in~(\ref{equ: recvSigVec_i}) into the \acrshort{fim} formula in~\cite[Eq.~(6.55)]{Schreier2010Statistical}, and the \acrshort{crlb} can be obtained based on~\cite[Eq.~(27)]{Elzanaty2021Reconfigurable}.
\end{IEEEproof}

Based on Proposition~\ref{prop: crlb expression}, we use the expected \acrshort{crlb} over the \acrshort{roi} to substitute the objective in (SP1). 
Moreover, to facilitate the calculation of expectation, we employ the Monte Carlo method to approximate the expected \acrshort{crlb} by the average \acrshort{crlb} for a set of \gls{numSamples} randomly sampled positions following distribution~\gls{userDistribu}, which is denoted by \gls{setSamples}.
Thus, (SP1) is converted into the following \acrshort{crlb} minimization problem:
\beq
\text{(SP1$'$):}\!\min_{\gls{sigSet}, \gls{codeSet}} \sum_{\gls{userPos} \in \gls{setSamples}} {\mathrm{CRLB}(\gls{userPos})\over \gls{numSamples}}, ~\text{s.t.}~ \text{(\ref{opt P1: cons 2})-(\ref{opt P1: cons 3})},~\forall i,j.
\eeq

\srev{\textbf{Remark 3}: Based on~\eqref{equ: cov mat of multipath} and Proposition~\ref{prop: crlb expression}, the impact of passive user interference on the CRLB can be analyzed: If the magnitudes of all entries of $\bm V_i$ increase by a factor of $X$ times due to larger passive user interference~($\forall i\in\gls{valueSet_i}$), then based on~\eqref{equ: general crlb expression}--\eqref{equ: some equ follow fim}, $\mathrm{CRLB}(\gls{userPos})$ will increase approximately by a factor of $X$ as well.}

\frev{\textbf{Remark 4}: According to Proposition~\ref{prop: crlb expression}, the benefits of using multiple bands for positioning are two-fold.
\emph{Firstly}, the Fisher information from multiple bands adds up, leading to a lower value for the CRLB.
\emph{Secondly}, the fact that signals received in different bands are generally less correlated (due to less correlated multipath gains) also contributes to a lower CRLB. This can be shown by deriving the Fisher information of two correlated signals based on~\cite[Eq. (3.31)]{kay1993fundamentals}.}

{\textbf{Remark 5}: Solving (SP1$'$) is robust to sparse channel condition.
Based on~\eqref{equ: fim matrix}, 
we can observe that the diagonal components of FIM are proportional to the magnitude of variation of the received signal vector w.r.t. the user position. 
Therefore, even in cases where the received signal only contain single component having all its energy, solving~(SP1$'$) with our proposed algorithm below can still achieve a small CRLB by enabling the received signals to change rapidly within the ROI.}

\subsubsection{Solving \acrshort{crlb} Minimization} 
\label{s3ec: crlb minimization alg}

In (SP1$'$), there are a large number of optimization variables in \gls{sigSet} and \gls{codeSet}. 
\revise{Besides, to ensure the objective function in (SP1$'$) approximates the actual expectation of the \acrshort{crlb} over the \acrshort{roi}, \gls{numSamples} also needs to be large. }
Consequently, evaluating the value and gradients of the objective function is highly computationally complex, which makes traditional optimization algorithms inefficient.
To solve (SP1$'$) efficiently, we design a proximal stochastic gradient descent algorithm based on the ProxSARAH algorithm in~\cite{Pham2020ProxSARAH}, which is a state-of-the-art proximal stochastic descent algorithm.

\revises{Similar to ProxSARAH, our algorithm involves an \emph{inner loop} and an \emph{outer loop}.}
\frev{In the outer loop, the gradient of the objective function w.r.t. the optimization variables is coarsely estimated for a set of sampling points in the ROI. 
Then, in the inner loop, the gradient estimate is iteratively refined based on the gradient deviation determined during the update of the optimization variables.
This method enhances the precision of gradient estimation for limited sampling points, leading to an accelerated convergence rate~\cite{Pham2020ProxSARAH}.}
Moreover, to efficiently handle the large number of optimization variables, we optimize \gls{sigSet} and \gls{codeSet} alternatingly.

The proposed algorithm is described as follows.
In each iteration of the outer loop, the inner loops for the \acrshort{dna} beamforming variables are conducted sequentially.
In the $\ell_{\outter}$-th iteration~($\ell_{\outter} = 1,\cpdots, N_{\outter}$) of the outer loop for instance, a \emph{gradient estimate} is first generated as
\beq
\label{equ: v0s}
\bm V^{(0)}_{\mathrm{S},i,j} = \frac{1}{|\mathcal B_{\mathrm{S}}^{(\ell_{\outter})}|}\sum_{\gls{userPos} \in \mathcal B_{\mathrm{S}}^{(\ell_{\outter})}}\nabla_{\bm S_{i,j}} \mathrm{CRLB}(\gls{userPos}; \bm S_{i,j}),~\forall i,j,
\eeq
where $\mathcal B_{\mathrm{S}}^{(\ell_{\outter})}$ denotes a randomly selected batch of position samples in \gls{setSamples}.

Then, the inner loop for digital beamforming starts with $\bm S_{i,j}^{(0)} = \gls{sigMat_ij}$. 
\revises{In the $\ell_{\inner}$-th iteration~($\ell_{\inner}=1,\cpdots,N_{\inner}$) of the inner loop}, the digital beamforming variables are updated as
\beq
\begin{aligned}
\label{equ: Sij inner update}
\bm S_{i,j}^{(\ell_{\inner})} =& (1-\gamma) \bm S_{i,j}^{(\ell_{\inner}-1)} \\
&+ \gamma \prox_{\mathrm{S}}(\bm S_{i,j}^{(\ell_{\inner}-1)} - \beta \bm V^{(\ell_{\inner}-1)}_{\mathrm{S},i,j}),~\forall i,j,
\end{aligned}
\eeq
where $\gamma$ and $\beta$ are coefficients controlling the step size of the update, and $\prox_\mathrm{S}(\cdot)$ denotes the proximal operator for the digital beamforming variables to ensure that constraint~(\ref{opt P1: cons 2}) is satisfied.
Then, the gradient estimate is iteratively updated in the inner loop as ($\forall i,j$)
\begin{align}
\label{equ: gradient estimate}
&\bm V^{(\ell_{\inner})}_{\mathrm{S},i,j} =  \bm V^{(\ell_{\inner} -1)}_{\mathrm{S},i,j} \\
&+ \hspace{-1em}\sum_{\gls{userPos} \in \mathcal B_{\mathrm{S}}^{(\ell_{\outter}, \ell_{\inner})}}\hspace{-1em}\frac{ \nabla_{\bm S_{i,j}} \mathrm{CRLB}(\gls{userPos}; \bm S_{i,j}^{(\ell_{\inner})}) - \nabla_{\bm S_{i,j}} \mathrm{CRLB}(\gls{userPos}; \bm S_{i,j}^{(\ell_{\inner}-1)})}{|\mathcal B_{\mathrm{S}}^{(\ell_{\outter}, \ell_{\inner})}|},\nonumber
\end{align}
where $\mathcal B_{\mathrm{S}}^{(\ell_{\outter}, \ell_{\inner})}$ is a randomly selected batch of samples in \gls{setSamples} to estimate the gradient difference.
The obtained $\bm V^{(\ell_{\inner})}_{\mathrm{S},i,j}$ is fed into~(\ref{equ: Sij inner update}) for the next iteration.
After \revises{$N_{\inner}$ iterations of the inner loop}, the current digital beamforming variables are updated as $\bm S_{i,j} = \bm S_{i,j}^{(N_{\inner})}$, $\forall i,j$.

Then, steps similar to~(\ref{equ: v0s})-(\ref{equ: gradient estimate})  are carried out for the analog beamforming variables, substituting symbols $\bm S_{i,j}$, $\bm V_{\mathrm{S},i,j}$, $\mathcal B_{\mathrm{S}}$, $\prox_{\mathrm{S}}(\cdot)$, and $\nabla_{\bm S_{i,j}} \mathrm{CRLB}(\gls{userPos}; \bm S_{i,j})$ with $\bm C_{i}$, $\bm V_{\mathrm{C},i}$, $\mathcal B_{\mathrm{C}}$, $\prox_{\mathrm{C}}(\cdot)$, and $\nabla_{\bm C_{i}} \mathrm{CRLB}(\gls{userPos}; \bm C_{i})$, respectively. 
Moreover, based on~\cite{Parikh2014Proximal}, in~(\ref{equ: Sij inner update}), the proximal operators for the \acrshort{dna} beamforming variables can be expressed as
\beq
\begin{aligned}
\label{equ: prox operator}
&[\prox_{\mathrm{S}}(\bm S_{i,j})]_{q} = \frac{[\bm S_{i,j}]_q}{\|[\bm S_{i,j}]_q\|_2} P_{\max}, \\
&[\prox_{\mathrm{C}}(\bm C_{i})]_{q,m} = 
\min(\max([\bm C_{i}]_{q,m},0), 1),
\end{aligned}
\eeq
where ~$ q\in\gls{valueSet_q}$ and $m\in\gls{valueSet_m}$.
Furthermore, we derive the gradients of the \acrshort{crlb} w.r.t. \gls{sigSet} and \gls{codeSet} in~(\ref{equ: v0s}) and~(\ref{equ: gradient estimate}) in close form in the following proposition.

\begin{proposition}
\label{prop: grad of crlb with s and c}
The gradient of the \acrshort{crlb} w.r.t. $\gls{sigMat_ij}$~($\forall i \in \gls{valueSet_i}, j\in \gls{valueSet_j}$) at position $\gls{userPos}$ can be calculated by the formulas as follows:
\begin{align}
\label{equ: gradient crlb wrt S-1}
&{\partial \mathrm{CRLB}(\gls{userPos}; \gls{sigMat_ij})\over \partial \gls{sigMat_ij}} = -  \trace\!\Big({\partial\gls{fimMat_userpos}\over \partial \gls{sigMat_ij}}\!\gls{inv2fimMat_userpos}\Big),\\
\label{equ: gradient crlb wrt S-2}
& \left[{\partial\gls{fimMat_userpos}\over \partial \gls{sigMat_ij}}\right]_{u,v} \hspace{-.5em} =\! 2\Re\Big({\bm A_{i,j,v,u}^{\mathrm s}} \!+\! {\bm A_{i,j,u,v}^{\mathrm s}} \!+\! {\bm B_{i,j,u,v}^{\mathrm s}} \!+\! {\bm B_{i,j,v,u}^{\mathrm s}}\Big)\!,\nonumber \\
&~\forall u,\!v \!\in\! \{1,\!2,\!3\}.
\end{align}
\revise{Here, notations $\bm A_{i,j,u,v}^{\mathrm s}$ and $\bm B_{i,j,u,v}^{\mathrm s}$ are defined in Appendix~\ref{appx: 1}.}
Besides, the gradient of the \acrshort{crlb} w.r.t. $\gls{codeMat_i}$~($\forall i \in \gls{valueSet_i}$) at $\gls{userPos}$ can be calculated in a similar manner as~(\ref{equ: gradient crlb wrt S-1}),~(\ref{equ: gradient crlb wrt S-2}) by substituting $\bm A_{i,j,u,v}^{\mathrm s}$, $\bm B^{\mathrm s}_{i,j,u,v}$, and \gls{sigMat_ij} with $\bm A_{i,u,v}^{\mathrm c}$, $\bm B^{\mathrm c}_{i,u,v}$, and \gls{codeMat_i}, respectively.
The complete expressions of $\bm A_{i,u,v}^{\mathrm c}$ and $\bm B^{\mathrm c}_{i,u,v}$ are also given in Appendix~\ref{appx: 1}.
\end{proposition}
\vspace{-.5em}
\begin{IEEEproof}
Please refer to Appendix~\ref{appx: 1}.
\end{IEEEproof}

In summary, the algorithm for \acrshort{dna} beamforming optimization is provided in Algorithm~\ref{alg: summary algorithm for dna beamforming}.
In the following subsections, $\{\bm S_{i,j}^{*}\}_{i,j}$ and $\{\bm C_i^{*}\}_i$ obtained by Algorithm~\ref{alg: summary algorithm for dna beamforming} are employed as default.

\begin{figure}[!t]
\vspace{-2em}
\begin{algorithm}[H]
\small
 \caption{\acrshort{dna} Beamforming Optimization Algorithm}
\label{alg: summary algorithm for dna beamforming}
\begin{algorithmic} [1]
\State Sample $\gls{numSamples}$ position samples following $\gls{userDistribu}$ and obtain $\gls{setSamples}$.
\State Set initial $\{\bm S_{i,j}^{(0)}\}_{i,j} = \{\gls{sigMat_ij}|s_{i,j,k}^{(q)}=\sqrt{\gls{userMaxPower}}/\gls{numFeed}\}$ and $\{\bm C_i^{(0)}\}_i$ containing random elements within $[0,1]$.
\For{$\ell_{\outter} = 1, \cpdots, N_{\outter}$}\quad \emph{\# Outer Loop}
\State Generate an initial gradient estimate for the \acrshort{crlb} with~(\ref{equ: v0s}), which is denoted by $\bm V^{(0)}_{\mathrm{S},i,j}$, $\forall i,j$.
\For{$\ell_{\inner} = 1, \cpdots, N_{\inner}$}\quad \emph{\# Inner Loop}
\State Update $\bm S_{i,j}^{(\ell_{\inner})}$ based on~(\ref{equ: Sij inner update}) and~(\ref{equ: prox operator}) by using gradient estimate $\bm V^{(\ell_{\inner} - 1)}_{\mathrm{S},i,j}$, $\forall i,j$.
\State Update the gradient estimate by (\ref{equ: gradient estimate}) with the help of the gradient formulas given in Proposition~\ref{prop: grad of crlb with s and c}.
\EndFor 
\State Conduct \revise{steps similar to} Steps 4 to 7 for analog beamforming variables \gls{codeSet}, $\forall i$.
\EndFor 
\State Return $\{\bm S_{i,j}^{*}\}_{i,j}$ and $\{\bm C_i^{*}\}_i$ as the current $\{\bm S_{i,j}\}_{i,j}$ and $\{\bm C_i\}_i$.
\end{algorithmic}
\end{algorithm}
\vspace{-1.5em}
\end{figure}

\subsection{\revise{Initial Point Selection and Adaptation Function Design}}
\label{s2ec: ini and ada}

Now, we focus on the second challenge of~(P1), i.e., the objective function cannot be effectively minimized as the user's local datasets only contain position labels for the areas around a few anchors.
To overcome this challenge, a proper adaptation function, $\{\bm{\mathcal A}^{(t)*}\}_t$, and a suitable initial parameter vector of the \acrshort{pos_est}, $\bm w^{(0)*}$, are needed.
Nevertheless, due to their high degrees of freedom, they are hard to obtain by traditional optimization algorithms.
To tackle this issue effectively, we employ the \emph{transfer learning} technique to obtain $\bm w^{(0)*}$ and $\{\bm{\mathcal A}^{(t)*}\}_t$.

To begin with, we describe the initial point selection and adaptation function design sub-problem of (P1) in the context of transfer learning as follows.
The target environment, where we aim to optimize \gls{SysName}, constitutes \emph{target domain} $\mathcal D_{\mathrm{tar}}=\{\gls{userDistribu}, \{\bm y_i(\cdot)\}_i\}$, which comprises the distribution of user position \gls{userDistribu} and the set of exact channel models $\{\bm y_i(\cdot)\}_i$.
Then, the objective of~(P1) can be considered as a task on $\mathcal D_{\mathrm{tar}}$ denoted by $\mathcal T(\mathcal D_{\mathrm{tar}})$, in which we aim to find the optimal parameter vector for the minimization of the positioning error given \gls{userDistribu} and $\{\bm y_i(\cdot)\}_i$.

Due to the undetermined environmental characteristics and the insufficient local datasets of the users, the solution to $\mathcal T(\mathcal D_{\mathrm{tar}})$ cannot be obtained by conventional optimization techniques.
Fortunately, the transfer learning technique provides an effective means to handle $\mathcal T(\mathcal D_{\mathrm{tar}})$.
Specifically, we resort to a domain similar to $\mathcal D_{\mathrm{tar}}$, where the task can be efficiently solved.
We refer to this domain as the \emph{source domain} and denote it by $\mathcal D_{\mathrm{src}}$, and the solution to $\mathcal T(\mathcal D_{\mathrm{src}})$ is denoted by $\bm w^{*}{}'$.
\srev{To obtain $\bm w^{*}{}'$, certain environmental characteristics need to be assumed for $\mathcal D_{\mathrm{src}}$, which are generally different from those in $\mathcal D_{\mathrm{tar}}$.
This results in different joint distributions for the user positions and received signals in the two domains, and hence $\bm w^{*}{}'$ is not valid in $\mathcal D_{\mathrm{tar}}$.}
\frev{Nevertheless, due to the intrinsic similarities between $\mathcal D_{\mathrm{tar}}$ and $\mathcal D_{\mathrm{src}}$ (e.g., the underlying signal propagation models and DA beamforming configurations are identical), transfer learning can be used efficiently to adapt $\bm w^{*}{}'$ to $\mathcal D_{\mathrm{tar}}$~\cite{Zhuang2021PIEEE_AComprehensize}.}
Therefore, $\mathcal T(\mathcal D_{\mathrm{tar}})$ can be handled by selecting $\bm w^{*}{}'$ as the initial point, i.e., $\bm w^{(0)}$, and adapting it to $\mathcal D_{\mathrm{tar}}$ with $\{\bm{\mathcal A}^{(t)}\}_t$, which is designed to minimize the positioning error of $\bm f(\cdot; \bm w^{(T)})$ over the users' local datasets.
In the following, we describe the detailed procedures for selecting the initial point and designing the adaptation function.

\subsubsection{\revise{Selection of Initial Point}}
\label{s3ec: sel ini p}
We choose the source domain having channel model $\{\tilde{\bm y}_i(\cdot)\}_i$ in Sec.~\ref{s2ec: source domain solving alg}, i.e., $\mathcal D_{\mathrm{src}} = \{ \gls{userDistribu}, \{\tilde{\bm y}_i(\cdot)\}_i\}$.
Then, by using $\{\tilde{\bm y}_i(\cdot)\}_i$, the \acrshort{bs} can generate sufficient received signal matrices with position labels.
Denote the generated dataset by $\mathcal S_{\mathrm{src}} = \{({\bm Y}_{\rx, \ell}, \bm p_{\ell})\}_{\ell}$ with $\ell \in\{ 1,\cpdots, \gls{numSimulDomData}\}$, where $\gls{numSimulDomData}$ is the size of $\mathcal S_{\mathrm{src}}$.
Then, $\mathcal T(\mathcal D_{\mathrm{src}})$ can be handled by solving the following optimization problem.
\beq
\label{opt: simulation domain learning}
\text{(SP2-1):}\min_{\bm w'}~\frac{1}{\gls{numSimulDomData}}\sum_{({\bm Y}_{\rx, \ell}, \bm p_{\ell})\in \mathcal S_{\mathrm{src}}}  \| \bm p_{\ell} - \bm f({\bm Y}_{\rx, \ell}; \bm w')\|_2^2.
\eeq
The solution to problem (SP2-1), ${\bm w'}^{*}$, can be obtained efficiently by using Adam algorithm~\cite{goodfellow2016deep}, and
we select $\bm w^{(0)*} = {\bm w'}^{*}$ as the proper initial point for the adaptation under the \acrshort{fl} framework.

\subsubsection{\revise{Design of Adaptation Function}}
\label{s3ec: design ada func}

Next, we design the adaptation function to adapt the solution to $\mathcal T(\mathcal D_{\mathrm{src}})$ to $\mathcal T(\mathcal D_{\mathrm{tar}})$, so that the objective function in~(P1) can be optimized.
Specifically, the adaptation function needs to satisfy two important conditions:
\emph{First}, the adaptation should not overfit the \acrshort{pos_est} to the limited local datasets of the users; otherwise, the resulting \acrshort{pos_est} may yield low positioning errors only around the anchors.
\emph{Second}, the adaptation should not be biased towards the local dataset(s) of one or few users; otherwise, the resulting \acrshort{pos_est} may only get low positioning errors for part of the users.

To satisfy the first condition while fully utilizing the limited target domain data, we employ a \emph{fine-tuning technique} to set different adaptation rates for different parts of parameter vector $\bm w^{(t)}$, where the \acrshort{mlp} of the \acrshort{pos_est} is viewed as being composed of two components: The output layer of the \acrshort{mlp} constitutes a \emph{regressor} deriving the user's position from the \emph{features} extracted by the other layers; and the other layers jointly constitute a \emph{feature extractor}.
We denote the parameters in $\bm w^{(t)}\in\mathbb R^{\gls{dimCoeff}}$ corresponding to the feature extractor and the regressor by $\bm w_{\mathrm{feat}}^{(t)}\in\mathbb R^{\gls{dimCoeff_feat}}$ and $\bm w_{\mathrm{reg}}^{(t)}\in\mathbb R^{\gls{dimCoeff_reg}}$, respectively, i.e., $\bm w^{(t)} = (\bm w_{\mathrm{feat}}^{(t)}, \bm w_{\mathrm{reg}}^{(t)})$ and $\gls{dimCoeff} = \gls{dimCoeff_feat} + \gls{dimCoeff_reg}$.

\frev{We note that the feature extractor optimized for $\mathcal T(\mathcal D_{\mathrm{src}})$ is also effective for $\mathcal T(\mathcal D_{\mathrm{tar}})$
as the channels in both domains follow the same structure,
i.e.,~(\ref{equ: gamma_ij})-(\ref{equ: recvSigVec_i}), and the same DA beamforming configurations are employed.
Therefore, the method for feature extraction needs little adaptation, and $\bm w_{\mathrm{feat}}^{(t)}$ can be frozen or adapted with a very low rate $\eta_{\mathrm{feat}}$.}
\srev{In contrast, the regressor has to be adapted substantially to handle the differences in the extracted feature values caused by the different environmental characteristics.
Thus, we adapt the regressor with a large learning rate denoted by $\eta_{\mathrm{reg}}$.
Though the amount of local user data collected in $\mathcal D_{\mathrm{tar}}$ is small, the adaptation of the regressor can still be done effectively since the regressor only contains the output layer of the MLP with a small number of trainable parameters.}
 
Besides, to satisfy the second condition, the aggregation function is expected to update the parameter vector along an unbiased gradient direction for minimization of the loss functions of all users, i.e.,
it should solve the following \emph{target domain adaptation optimization} problem:
\begin{align*}
(\text{SP2-2}): \min_{\{\bm {\mathcal A}^{(t)}\}_{t}} & \hat{\mathcal L}(\bm w^{(T)}) = \sum_{n=1}^{\gls{numUser}} \hat{\mathcal L}_n(\bm w^{(T)}),\\
\text{s.t.}~&\text{(\ref{opt P1: cons 4})-(\ref{opt P1: cons 5})},~\bm w^{(0)} = \bm w^*_{\mathrm{src}},~\forall t=1,\cpdots, T,
\end{align*}
where $\hat{\mathcal L}_n(\bm w^{(T)})$ is defined in~(\ref{equ: def hat L n}) and $\hat{\mathcal L}(\bm w^{(T)})$ denotes the total loss of all users w.r.t. $\bm w^{(T)}$.
Therefore, in each positioning process $t$, the selected adaptation function $\bm{\mathcal A}^{(t)*}$ should update $\bm w^{(t)}$ in the opposite direction of an unbiased estimate of $\nabla_{\bm w}\hat{\mathcal L}(\bm w^{(t)})$, which is denoted by $\bm g^{(t)}$.

To obtain this unbiased estimate, based on~\cite[Lemma~1]{Ren2020Scheduling}, we can multiply the uploaded local update from user $x$, i.e., $\Delta \bm w_x^{(t)}$, with a \emph{weight} which is in proportion to the size of user $x$'s local dataset, i.e., $Q_x$, and in inverse proportion to its scheduling probability, i.e., $\xi_{x}^{(t)}$, so that 
\beq
\label{equ: def of gt}
\mathop{\mathbb E}\limits_{x\sim\mathcal M(\bm \xi^{(t)})} \Big(\frac{Q_x}{Q \xi^{(t)}_x} \Delta \bm w_x^{(t)}\Big)=  -\nabla_{\bm w}\hat{\mathcal L}(\bm w^{(t)}) =  - \bm g^{(t)},
\eeq
where $Q = \sum_{n=1}^{\gls{numUser}} Q_n$ is the total size of all users' local datasets.
In summary, based on the fine-tuning technique and~(\ref{equ: def of gt}), the set of adaptation functions to solve (SP2-2) can be designed as
\beq
\label{equ: designed A t}
\bm{\mathcal A}^{(t)*}(\bm w^{(t)}\!, \Delta \bm w_x^{(t)}) \!=\! \bm w^{(t)} + \bm \eta \odot  \Big(\frac{Q_x}{Q \xi^{(t)}_x} \Delta \bm w_x^{(t)}\Big),
\eeq
$\forall t= 1,\cpdots, T$, where $\bm \eta = (\eta_{\mathrm{feat}}\bm 1_{\gls{dimCoeff_feat}}, \eta_{\mathrm{reg}}\bm 1_{\gls{dimCoeff_reg}})$ is the adaptation rate vector.

\subsection{User Scheduling Probability Optimization}
\label{s2ec: alg design ftl}

In the following, we handle the third challenge of (P1) by optimizing the user scheduling probability in each positioning process.
Specifically, when optimizing $\bm \xi^{(t)}$, we consider two important factors as follows.
\emph{First}, in order to achieve fast convergence in FL, we consider the impact of $\bm \xi^{(t)}$ on the expected convergence rate. 
\emph{Second}, to efficiently utilize the spectrum resource, we also evaluate the effect of $\bm\xi^{(t)}$ on the efficiency of gradient uploading.

Here, the convergence rate can be analyzed by extending~\cite[Lemma~2]{Ren2020Scheduling} to the \acrshort{fl} framework under the proposed federated positioning protocol, as shown below.
\begin{proposition}
\label{prop: converge rate}
Given $\bm{\mathcal A}^{(t)*}(\cdot)$ in (\ref{equ: designed A t}), denote the optimal parameter vector for (SP2-2) by $\bm w^*$. 
The expected convergence rate for the $t$-th positioning process is characterized by
\begin{align}
\label{equ: converge rate}
 \mathbb E &\big(\hat{\mathcal L}(\bm w^{(t+1)}) \!- \! \hat{\mathcal L}(\bm w^{*})\big) \nonumber\\
& \leq \mathbb E\big(\hat{\mathcal L}(\bm w^{(t)}) \!-\! \hat{\mathcal L}(\bm w^{*})\big) - \gls{lrVec}^{\tran}\Big( \bm 1-\frac{L}{2} \gls{lrVec}\Big) \odot \bm{g}^{(t)\circ 2} \nonumber \\[-.5em]
& + \frac{L}{2} 
   \sum_{n=1}^{\gls{numUser}} {1\over \xi_n^{(t)}}  \left(\frac{Q_n}{Q}\right)^2  \big(\gls{lrVec}^{\circ 2 \tran}\mathbb E(\bm g_n^{(t)\circ 2}) + \gls{dpSigma_n2}\|\gls{lrVec}\|^2\big)  \nonumber \\
& - \frac{L}{2} \gls{lrVec}^{\circ 2 \tran}\bm{g}^{(t)\circ 2}.
\end{align}
\revise{Here, $\bm g^{(t)}$ and $\bm g_n^{(t)}$ are the gradients of $\hat{\mathcal L}(\bm w^{(t)})$ and $\hat{\mathcal L}_{n}(\bm w^{(t)})$, respectively, as defined in (\ref{equ: def of gt}) and below~(\ref{equ: def hat L n}), $L$ is the Lipschitz constant of the gradient, and $\gls{lrVec}$ is defined in~(\ref{equ: designed A t}).}
\end{proposition}
\vspace{-.5em}
\begin{IEEEproof}
Please refer to Appendix~\ref{appx: proof of prop converge rate}.
\end{IEEEproof}

In~(\ref{equ: converge rate}), the only term related to $\bm \xi^{(t)}$ is the third term on the right-hand side of the inequality, which reflects the influence of $\bm \xi^{(t)}$ on the convergence rate and needs to be minimized. 
\frev{Proposition~\ref{prop: converge rate} reveals that the convergence rate is dependent on the powers of the gradients of the users and, to improve the rate of convergence, users having higher gradient powers should be scheduled with higher probabilities.}
However, as $\mathbb E(\bm g_n^{(t)\circ 2})$ is difficult to estimate by the \acrshort{bs} or the users, we approximate it by $\bm g_n^{(t)\circ 2}$ as in~\cite{Ren2020Scheduling}.
Based on~(\ref{equ: converge rate}), for the scheduling probability of user~$n$ uploading in positioning process $t$, i.e., $\xi^{(t)}_n$, we define its \emph{influence on convergence} as
\beq
\label{equ: Z cr x}
Z_{\mathrm{IC},n}^{(t)} = \left(\frac{Q_n}{Q}\right)^2 \cdot \Big(\gls{lrVec}^{\circ 2 \tran}\mathbb E(\bm g_n^{(t)\circ 2}) + \gls{dpSigma_n2} \|\gls{lrVec}\|^2\Big).
\eeq

Besides, we evaluate the influence of $\xi^{(t)}_n$ on the uploading efficiency by the ratio between \emph{uploading duration} and \emph{weighted uploading capacity}.
The uploading duration is the time duration for the uplink transmission of user $n$'s local update, and the uploading capacity is the maximum information on the optimization step of $\bm w^{(t)}$ contained in the uploaded gradient.
We measure the uploading capacity based on the Shannon channel capacity formula\cite{goldsmith2005wireless}.
Intuitively, the square of the uploaded gradient can be considered as the ``transmit power'', as a gradient with a larger squared norm can contain more information regarding the optimization step. The variance of the noise added to the gradient for \acrshort{dp} training, i.e., $\gls{dpSigma_n2}$,  can be considered as the noise power of the ``channel''.
Therefore, adding the weights for the learning rate and the data size, we define the \emph{weighted uploading capacity} of user $n$'s uploading in positioning process $t$ as
\beq
\Xi_n^{(t)} = {Q_n\over Q}\cdot\log\Big(1+\frac{\gls{lrVec}^{\circ 2 \tran}\mathbb E(\bm g_n^{(t)\circ 2})}{\gls{dpSigma_n2} \|\gls{lrVec}\|_2^2}\Big).
\eeq
The \emph{influence on uploading efficiency} of $\xi^{(t)}_n$ is calculated as
\beq
\label{equ: Z eu x} 
Z_{\mathrm{IE},n}^{(t)} = \frac{\gls{sizeCoeff}}{R_{n}^{(t)}\Xi_n^{(t)}},
\eeq
where \gls{sizeCoeff} denotes the total size of the gradient vector in bits, and $R_{n}^{(t)}$ denotes the uplink transmission rate from user $n$ to the \gls{bs}\footnote{%
\srev{As user $n$ can calculate $Z_{\mathrm{IE},n}^{(t)}$ in \eqref{equ: Z eu x} with $R_n^{(t)}$ being the uplink rate of the user, the proposed user scheduling algorithm can be extended to arbitrary band selection schemes for uplink transmission of local gradients.}%
}.
Here, exploiting channel reciprocity, $R_{n}^{(t)}$ can be obtained by user $n$ based on the downlink frames from the \gls{bs} for distributing $\bm w^{(t)}$.

To allow the \acrshort{bs} getting enough information to optimize $\bm \xi^{(t)}$ while minimizing the overheads, during the federated adaptation phase, let each user $n$ calculate $Z_{\mathrm{IC},n}^{(t)}$ and $Z_{\mathrm{IE},n}^{(t)}$ based on its local gradient and send these values to the \acrshort{bs}.
Since $Z_{\mathrm{IC},n}^{(t)}$ and $Z_{\mathrm{IE},n}^{(t)}$ are scalars and contain only the norms of the local gradients, the cost of uplink transmission and privacy leakage is negligible.

Jointly considering the convergence rate and uploading efficiency, we formulate a \emph{joint convergence and efficiency scheduling optimization problem} as follows:
\begin{align*}
(\text{SP3}): \min_{\bm \xi^{(t)}}~&\sum_{n=1}^{\gls{numUser}}\Big(\frac{1}{\xi_{n}^{(t)}} \cdot \frac{1}{\hat{Z}_{\mathrm{IC}}^{(t)}}\cdot Z_{\mathrm{IC},n}^{(t)}  +\xi_{n}^{(t)}\cdot \frac{1}{\hat{Z}_{\mathrm{IE}}^{(t)}}\cdot Z_{\mathrm{IE},n}^{(t)}\Big),\\
\text{s.t.}~&\text{(\ref{opt P1: cons 6})}.
\end{align*}

In~(SP3), we employ two weight coefficients, i.e., $\hat{Z}_{\mathrm{IC}}^{(t)}$ and $\hat{Z}_{\mathrm{IE}}^{(t)}$, to rescale the impact of the convergence rate and the uploading efficiency. 
By this means, we prevent the objective of~(SP3) from being overly biased to either the convergence rate or the uploading efficiency given their different value ranges.
Specifically, $\hat{Z}_{\mathrm{IC}}^{(t)}$ and $\hat{Z}_{\mathrm{IE}}^{(t)}$ are set as follows:
\beq
\hat{Z}_{\mathrm{IC}}^{(t)} = \sum_{n=1}^{\gls{numUser}}\!\frac{Z_{\mathrm{IC},n}^{(t)}}{(\xi_{n}^{(t-1)})^2},\hat{Z}_{\mathrm{IE}}^{(t)} =\! \sum_{n=1}^{\gls{numUser}}Z^{(t)}_{\mathrm{IE},n}, \forall t\!=\!\{1,\cpdots, T\},
\eeq
with $\xi_n^{(0)} = 1/\gls{numUser}$.
It is straightforward to show that~(SP3) is convex and has a unique optimal solution, which can be efficiently found by using convex optimization algorithms.

\begin{figure}[!t]
\vspace{-2em}
\begin{algorithm}[H]
\small
\caption{Positioning Error Minimization Algorithm for \gls{SysName}}
\label{alg: ftl framework}
\begin{algorithmic} [1]
\State Obtain \gls{sigSet_opt} and \gls{codeSet_opt} with Algorithm~\ref{alg: summary algorithm for dna beamforming}.
\State Given \gls{sigSet_opt} and \gls{codeSet_opt}, generate the training data in the source domain, i.e., $\mathcal D_{\mathrm{src}}$.
\State Solve problem~(SP2-1) with the Adam algorithm~\cite{goodfellow2016deep} and obtain ${\bm w'}^{*}$ as $\bm w^{(0)}$.
\For{$t = 0,...,T$}
\State The \gls{bs} distributes $\bm w^{(t)}$ to all the users, and each user $n$ determines its uplink rate $R_n^{(t)}$ to the \gls{bs}.
\State Each user $n$ calculates its local gradient by $\bm g_n^{(t)} = \nabla_{\bm w} \hat{\mathcal L}_n(\bm w^{(t)})$.
\State \revise{Each user $n$ calculates $Z_{\mathrm{IC},n}^{(t)}$ and $Z_{\mathrm{IE},n}^{(t)}$ based on~(\ref{equ: Z cr x}) and (\ref{equ: Z eu x}) and sends them to the \gls{bs}.}
\State The \gls{bs} obtains the optimized scheduling probabilities, i.e., $\bm \xi^{(t)*}$, by solving~(SP3).
\State The \gls{bs} randomly selects user $x$ to upload its local update according to distribution $\mathcal M(\bm \xi^{(t)*})$.
\State \revise{User $x$ sends local update $\Delta \bm w_n^{(t)} = - \bm g_n^{(t)} + \bm \varsigma_n^{(t)}$ to the \gls{bs} with $\bm \varsigma_n^{(t)}$ being the noise for differential privacy.}
\State The BS updates the global parameter vector as $\bm w^{(t+1)} = \bm{\mathcal A}^{(t)*}(\bm w^{(t)}, \Delta \bm w_x^{(t)})$ based on~(\ref{equ: designed A t}).
\EndFor
\end{algorithmic}
\end{algorithm}
\vspace{-1.5em}
\end{figure}

\revise{Finally, integrating the \acrshort{dna} beamforming optimization, the initial point selection and adaptation function design, and the user scheduling probability optimization, the complete positioning error minimization algorithm for \gls{SysName} is summarized as Algorithm~\ref{alg: ftl framework}.}

\section{Simulation Results}
\label{sec: evaluation}

In this section, the simulation is described and then our key simulation results are provided.

\subsection{Simulation Setup}
\label{s2ec: simulation setup}

We establish a 3D coordinate system with its origin at the center of the \acrshort{rhs}, its x-axis along the perpendicular direction of the \acrshort{rhs}, and its z-axis pointing vertically upward.
\acrshort{roi} $\mathcal P$ is a cuboid region with its center at $(10,0,0)$~m and 3D dimensions $(\gls{roi_sideLen_x},\gls{roi_sideLen_y}, \gls{roi_sideLen_z}) = (10,10,2)$~m.
The distribution of the user positions, $\gls{userDistribu}$, is a 3D uniform distribution within $\mathcal P$.

The \acrshort{rhs} board is made of FR-4, which is a typical dielectric material used for printed circuit boards and has refractive index $n_{\mathrm{r}}=2.1$.
The \gls{numBand} bands of the system are centered at $f_i = (2+0.5i)$~GHz ($i\in\gls{valueSet_i}$) and the average wavelength of the center frequencies is denoted by $\lambda_{\mathrm{avr}}$.
The spacing between adjacent \acrshort{me}s is set to be $0.3\lambda_{\mathrm{avr}}$.

As for \gls{pos_est} $\bm f(\cdot;\bm w)$, a 5-layered \acrshort{mlp} with $(1024,512,128)$ nodes in the three hidden layers is employed.
Specifically, the output layer is treated as the regressor and the other layers are treated as the feature extractor, whose learning rates are set to $\eta_{\mathrm{reg}}=10^{-3}$ and $\eta_{\mathrm{feat}}=10^{-6}$, respectively.
We consider the case with $\gls{numUser} = 2$ users, where each user has $Q_n = 200$ labeled data in its local dataset, as default.
\Copy{R1-2-1}{\frev{Each data-label pair comprises a received signal matrix and the corresponding position label.
A user obtains a position label when it is within $0.25$~m of one of $10$ anchors. For simulation, the positions of the $10$ anchors are drawn from a uniform distribution within the \acrshort{roi}.}}
The other default parameters are listed in Table~\ref{table: parameter}.

\begin{table}
\centering
\caption{Simulation Parameters}
\label{table: parameter}
\vspace{-.5em}
\renewcommand{\arraystretch}{1.2}
\begin{scriptsize}
\begin{tabular}{||l|l||l|l||}
\hline
\textbf{Parameter}     & \textbf{Value}     & \textbf{Parameter}            & \textbf{Value}    \\ \hline
\gls{numBand}           & $2$               & \gls{numElem}                 & $10\times 10$      \\ \hline
\gls{numFeed}           & $3$               & \gls{rmsValue_i}               & $0.5~\mu$s         \\ \hline
\gls{numSBand}          & $8$               & \gls{userMaxSpeed}            & $20$ km/h          \\ \hline
\gls{durFrame}           & $4~\mu$s         & \gls{noisePowerDensity}        & $-174$ dBm/Hz      \\ \hline
\gls{numFrame}           & $4$              & \gls{userMaxPower}             & $1$ mW              \\ \hline
\gls{subBandwidth}      & $125$ kHz         & \gls{numSamples}               & $10^{4}$            \\ \hline
$T$                      & $400$            & \gls{numUser}                  & $2$                 \\ \hline
\gls{numSimulDomData}   & $10^5$            & $|\mathcal B_{\mathrm{S}}|, |\mathcal B_{\mathrm{C}}|$  & $40$    \\ \hline
$N_{\outter}$           & $500$             & $N_{\inner}$                   & $5$                 \\ \hline
$\gamma$                & $0.95$            & $\beta$                         & $0.1$               \\ \hline
\gls{dpDelta}           & $10^{-5}$         & \gls{sizeCoeff}                & $2.7\times 10^{7}$ bits \\ \hline
\end{tabular}
\end{scriptsize}
\vspace{-1.5em}
\end{table}

Besides, for the channel model in the source domain, the gains of each \acrshort{me} and user antenna are normalized as $g_{i,j}^{\elem}(\cdot)=g_{i,j}^{\user}(\cdot)=1$~($\forall i,j$).
Based on~\cite{Barriac2006Space}, we set $P_{\mathrm{pap},i}(\bm \theta)$ in~(\ref{equ: cov mat of multipath}) as a Laplacian function with zero mean and angular spread $10^{\circ}$ in both azimuth and elevation, scaled by the average LoS power within the \acrshort{roi}.
As for the exact channel model in the target domain, we assume a different gain pattern for the \acrshort{me}s, i.e., $g_{i,j}^{\elem}(\bm \theta) = \cos^{0.1}(\bm \theta)$ and different multipath characteristics, i.e., $P_{\mathrm{pap},i}(\bm \theta)$ with mean angle $10^\circ$ and angular spread $15^\circ$.

\subsection{Results and Analyses}
\label{s2ec: simulation result and analysis}
\begin{figure}[t]
\centering
\includegraphics[width=1\linewidth]{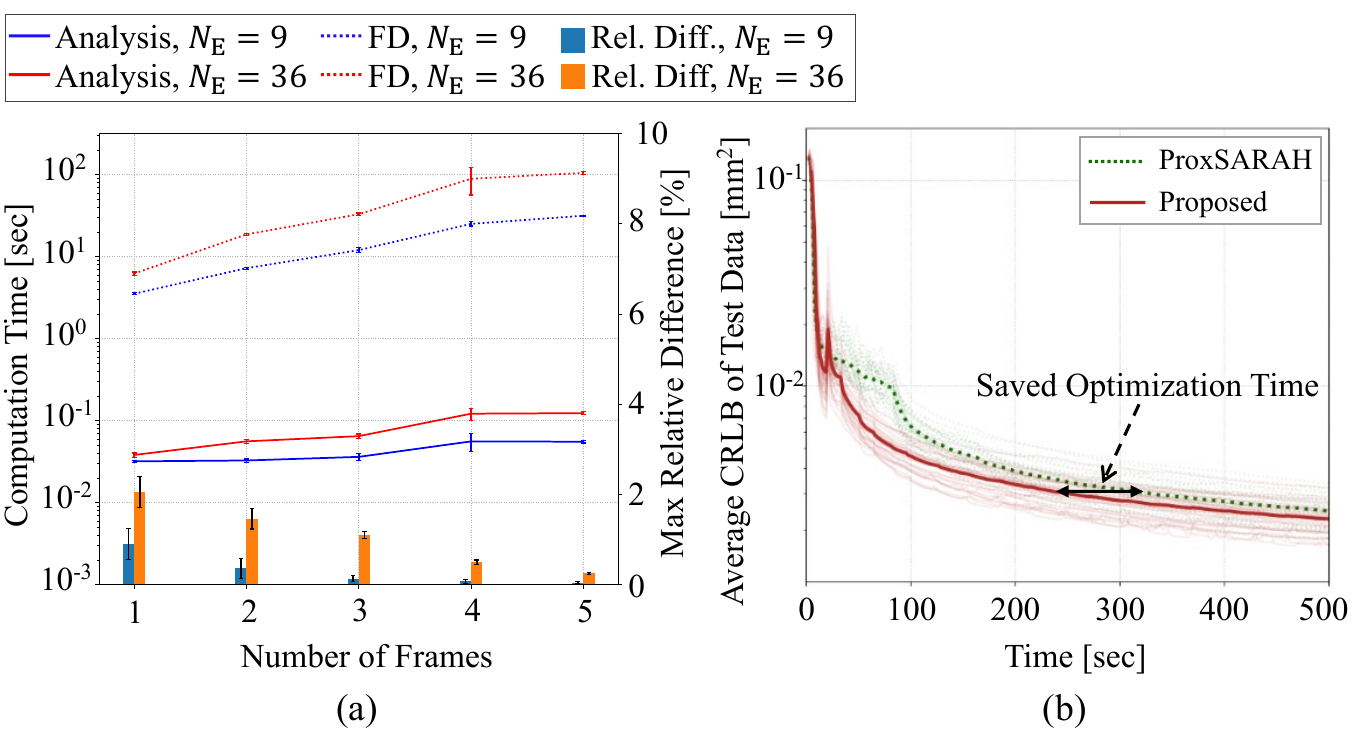}
  \caption{(a) Computational time of the gradient formulas in Proposition~\ref{prop: grad of crlb with s and c}~(\emph{Analysis}) and \revise{the finite difference method provided by MATLAB\textsuperscript{\textregistered}}~(\emph{\acrshort{fd}})  and their maximum relative difference~(\emph{Rel. Diff.}). The lines show the computational time, and the bars show the maximum element-wise relative difference. The error bars indicate the standard deviation over $30$ independent trials. (b) Comparison between the proposed \acrshort{dna} beamforming optimization algorithm and the benchmark ProxSARAH algorithm in~\cite{Pham2020ProxSARAH}. The translucent and opaque lines show results for $30$ individual trials and the average results, respectively.}
  \label{fig: simul fig 1}
\end{figure}

\emph{Firstly}, we validate the \acrshort{crlb} gradient formulas in Proposition~\ref{prop: grad of crlb with s and c}.
Fig.~\ref{fig: simul fig 1}~(a) shows the computational time of the proposed formulas and the finite difference~(\acrshort{fd}) method~\cite{Milne2000Calculus} and the maximum relative difference between them, which is calculated by $\max_{\ell}\{|[\tilde{\bm x}]_\ell \!-\![\bm x]_\ell|\!/\!\max(|[\bm x]_\ell|,\!1)\!\}$. 
Here, $\tilde{\bm x}$ and $\bm x$ denote the gradient vectors calculated by the proposed formulas and the \acrshort{fd} method, respectively.
The \acrshort{fd} method is implemented based on the function $\mathrm{finitedifferences}$ provided by MATLAB\textsuperscript{\textregistered}.
As can be observed in~Fig.~\ref{fig: simul fig 1}, the proposed formulas accurately determine the \acrshort{crlb} gradient with a significantly smaller computational time compared to the \acrshort{fd} method.
The maximum relative difference decreases with \gls{numFrame} since the \acrshort{crlb} decreases with \gls{numFrame}, and it increases with \gls{numElem} as a larger number of variables incurs more numerical errors. 

\emph{Secondly}, we verify the efficiency of the proposed \acrshort{dna} beamforming optimization algorithm in terms of \acrshort{crlb} minimization.
We compare it with the benchmark ProxSARAH algorithm~\cite{Pham2020ProxSARAH} through $30$ independent trials.
In each trial, $10^4$ random user positions within the \acrshort{roi} are sampled, where $90\%$ of them are used for optimization and $10\%$ of them are used as test data to evaluate the average \acrshort{crlb}.
In Fig.~\ref{fig: simul fig 1}~(b), it can be observed that, for the proposed algorithm, the average \acrshort{crlb} decreases with a faster rate than for ProxSARAH.
This indicates that the proposed algorithm is more efficient due to the alternation between the \acrshort{dna} variables.
On average, the proposed algorithm saves $37.5\%$ optimization time.

\emph{Thirdly}, we verify the effectiveness of the proposed \acrshort{dna} beamforming optimization in terms of the resulting \acrshort{mse} of positioning in the source domain.
We compare the \acrshort{dna} beamforming configuration optimized by Algorithm~\ref{alg: summary algorithm for dna beamforming} with two baseline beamforming configurations from~\cite{Zohair2021Near}:
\begin{itemize}[leftmargin=*]
\item \textbf{Directional Beams}~(\emph{DireBeam}): The \acrshort{dna} beamforming configuration generates focused beams scanning the \acrshort{roi} during the frames in the \acrshort{mmht} phase in Sec.~\ref{s3ec: MMHT phase}.
\item \textbf{Random Beams}~(\emph{RandBeam}): The elements of \gls{codeSet} are randomly distributed within $[0,1]$, and the elements of \gls{sigSet} take uniform values while satisfying the power constraint in~(\ref{opt P1: cons 2}).
\end{itemize}

\begin{figure}[t]
\centering
\includegraphics[width=0.6\linewidth]{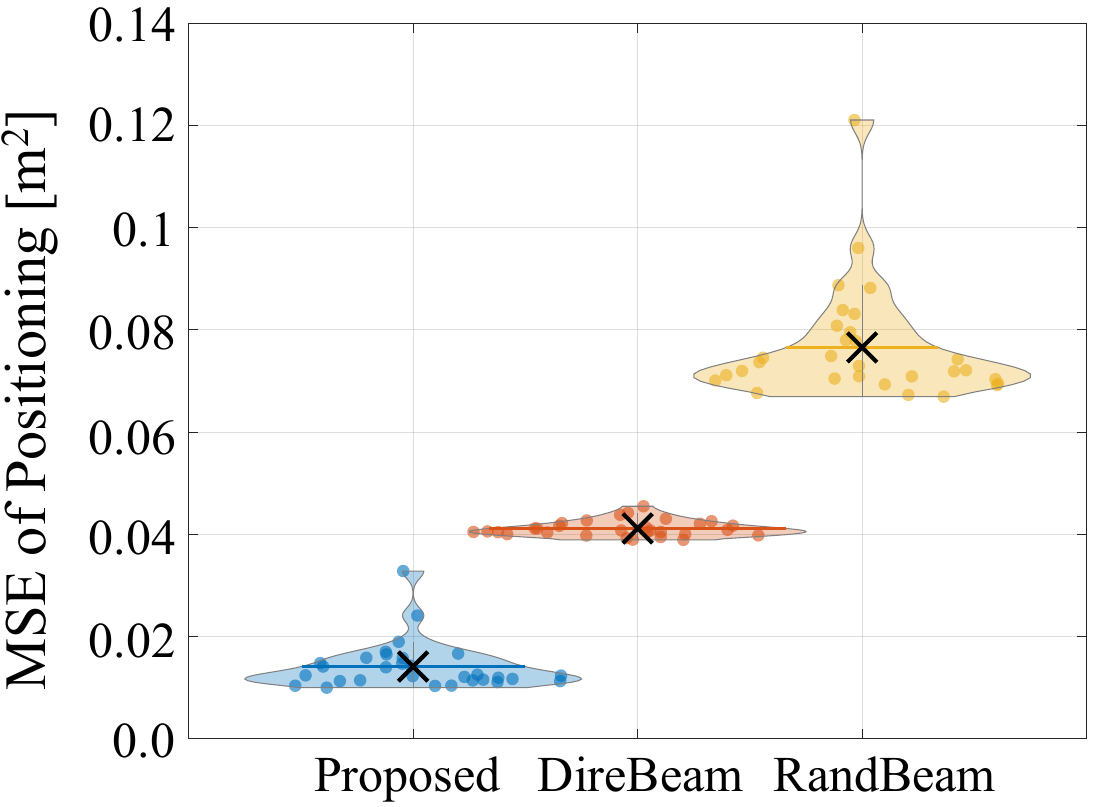} 
\caption{Comparison of the \acrshort{mse} of positioning for different \acrshort{dna} beamforming configurations.} 
\label{fig: simul fig 3}
\end{figure}

Fig.~\ref{fig: simul fig 3} shows the violin plot comparing the performance of different \acrshort{dna} beamforming configurations in terms of the resulting \acrshort{mse} of positioning.
The \acrshort{mse} of positioning is evaluated by the \gls{pos_est} optimized for the source domain by solving~(SP2-1) for each \acrshort{dna} beamforming configuration.
\Copy{R1-9-1}{\frev{To reduce randomness, for each configuration, we evaluate the \gls{pos_est} for $30$ randomly generated test sets, and the resulting average MSE of positioning in each trial is shown by a dot in Fig.~\ref{fig: simul fig 3}.}}
As can be observed from Fig.~\ref{fig: simul fig 3}, by using the \acrshort{dna} beamforming configuration obtained with Algorithm~\ref{alg: summary algorithm for dna beamforming}, \gls{SysName} can reduce the \acrshort{mse} of positioning by $57\%$ and $78\%$ on average compared to the DireBeam and RandBeam baselines, respectively.

\begin{figure}[!b]
\centering
\includegraphics[width=1\linewidth]{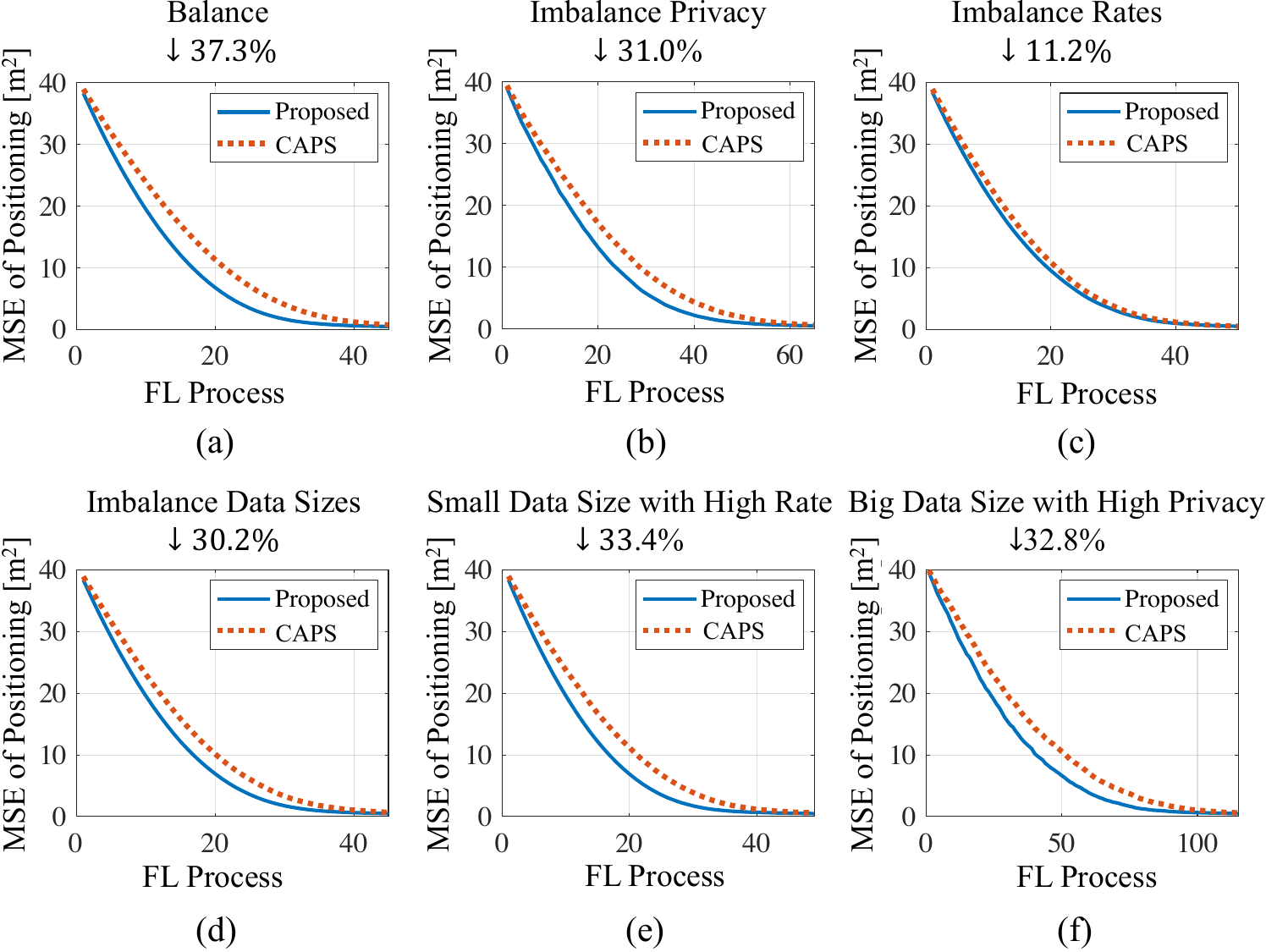}
  \caption{Comparison of the adaptation performance in \acrshort{fl} for the proposed user scheduling and the benchmark CAPS algorithms in $6$ typical situations: (a) Users have balanced privacy leakage bounds, uplink rates, and local data sizes;  (b) Users have imbalanced privacy leakage bounds, i.e., $(100, 20)$; (c) Users have imbalanced uplink rates, i.e., $(1,0.2)$ Mbps; (d) Users have imbalanced local data sizes, i.e., $(100,20)$; (e)~The user with a smaller data size has a higher uplink rate, i.e., rates $(0.2, 1)$ Mbps and data sizes $(100,20)$; (f)~The user with a larger data size has a tighter privacy leakage bound, i.e., local data size $(200,100)$ and privacy leakage bounds $(5,100)$.}
  \label{fig: simul fig 4}
\end{figure}

\emph{Fourthly}, we verify the efficiency of the proposed algorithm for user scheduling probability optimization in Sec.~\ref{s2ec: alg design ftl}.
We compare the results of the proposed algorithm with a state-of-the-art benchmark proposed in~\cite{Ren2020Scheduling}, which is referred to as channel-aware probabilistic scheduling~(CAPS) algorithm.
To facilitate a meaningful comparison, \revise{we compare the two scheduling algorithms for $6$ typical situations}, which are described in the caption of Figs.~\ref{fig: simul fig 4}(a)-(f).
Besides, to make the impact of scheduling on \acrshort{fl} as prominent as possible, the parameters of the regressors in the \acrshort{mlp}s are re-initialized randomly before the first positioning process.

Figs.~\ref{fig: simul fig 4}(a)-(f) show the adaptation  performance in terms of the \acrshort{mse} of positioning in different positioning processes, averaged over $30$ independent trials.
The proposed algorithm outperforms the benchmark algorithm in all considered situations.
\srev{Moreover, we calculate the average relative gain in terms of the required number of epochs for the \acrshort{mse} of positioning to drop from its initial value to below $0.5$ m$^2$.
The corresponding values are provided in the sub-headings of Figs.~\ref{fig: simul fig 4}(a)-(f), e.g., $\downarrow 37.3\%$ in Fig.~\ref{fig: simul fig 4}(a).}
It can be observed that the gains of the proposed algorithm are over $30\%$ except for the imbalance rate situation. 
This verifies that the proposed algorithm is more efficient than the benchmark algorithm in terms of training the position estimator to adapt to the target domain.

\begin{figure}[t]
\centering
\includegraphics[width=0.95\linewidth]{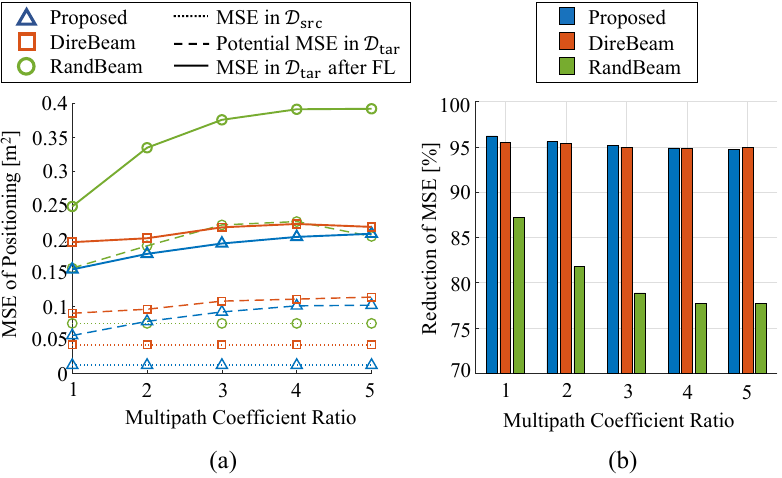}
\vspace{-.3em}
  \caption{\revise{(a) Comparison between the \acrshort{mse}s of positioning for different \acrshort{dna} beamforming configurations. (b) Comparison between the reduction of the \acrshort{mse} due to the adaptation in \acrshort{fl} for different \acrshort{dna} beamforming configurations.}}
  \vspace{-1.5em}
  \label{fig: simul fig 5}
\end{figure}

\emph{Fifthly}, we compare the \acrshort{mse} of positioning for the \acrshort{dna} beamforming configurations obtained by Algorithm~\ref{alg: summary algorithm for dna beamforming} and the DireBeam and RandBeam baselines for different levels of deviation between the source and target domains.
We control the deviation by changing the multipath coefficient ratio from $1$ to $5$, which represents the ratio between the angular spread of $P_{\mathrm{pap},i}(\bm \theta)$ for the target domain and that for the source domain.
The comparison focuses on three values: 1) the MSE in $\mathcal D_{\mathrm{src}}$, 2) the potential MSE in $\mathcal D_{\mathrm{tar}}$, and 3) the MSE in $\mathcal D_{\mathrm{tar}}$ after \acrshort{fl}.
Specifically, we evaluate the MSE in $\mathcal D_{\mathrm{src}}$ and the potential MSE in $\mathcal D_{\mathrm{tar}}$ by using supervised learning to train the \acrshort{pos_est} with $10^{5}$ labeled data in $\mathcal D_{\mathrm{src}}$ and $\mathcal D_{\mathrm{tar}}$, respectively; and we evaluate the \acrshort{mse} in $\mathcal D_{\mathrm{tar}}$ after \acrshort{fl} by using the \acrshort{pos_est} adapted to $\mathcal D_{\mathrm{tar}}$ with Algorithm~\ref{alg: ftl framework}.
Fig.~\ref{fig: simul fig 5}~(a) verifies that the \acrshort{dna} beamforming configuration obtained with the proposed algorithm leads to the lowest potential \acrshort{mse} in $\mathcal D_{\mathrm{tar}}$ and the lowest \acrshort{mse} in $\mathcal D_{\mathrm{tar}}$ after \acrshort{fl}.

\emph{Finally}, we verify the performance of \gls{SysName} in adapting to diverse environments by showing that the \acrshort{mse} of positioning in different target domains can be effectively reduced through the adaptation in the proposed protocol.
Fig.~\ref{fig: simul fig 5}~(b) shows the reduction of the \acrshort{mse} of positioning \emph{before} and \emph{after} the adaptation.
It can be observed that for different levels of deviation between the target and source domains, the federated adaptation consistently reduces the \acrshort{mse} of positioning by around $95\%$ for both the proposed and the DireBeam \acrshort{dna} beamforming configurations.

\section{Conclusion}
\label{sec: conclu}

In this paper, we have proposed \gls{SysName}, a user positioning system based on \acrshort{mb}-\acrshort{rhs} and \acrshort{fl} which can adapt to diverse practical environments.
We have formulated a positioning error minimization problem for \gls{SysName} and solved it by decomposing the problem into three subproblems.
First, we derived the \acrshort{crlb} of the positioning error and utilized it for optimization of the \acrshort{dna} beamforming configuration of the \acrshort{rhs}.
Second, we exploited transfer learning to select the initial point and adaptation function in \acrshort{fl}.
Third, we proposed a user scheduling probability optimization algorithm, jointly considering the convergence rate and uploading efficiency of \acrshort{fl}.
Simulation results have shown that the proposed \acrshort{dna} beamforming optimization algorithm can reduce the computation time required by ProxSARAH by $37.5\%$ and result in a $57\%$ lower \acrshort{mse} of positioning compared to DireBeam baseline.
Moreover, the proposed user scheduling optimization algorithm achieves a $11\%\sim 37\%$ lower average \acrshort{mse} in \acrshort{fl} process compared to the benchmark CAPS.
Furthermore, we showed that \gls{SysName} can adapt to diverse environments via federated adaptation, which can reduce the \acrshort{mse} of positioning by around $95\%$.

\setlength{\abovedisplayskip}{4pt}
\setlength{\belowdisplayskip}{4.5pt}

\begin{appendices}
\section{Components of The Gradients of The CRLB}
\label{appx: 1}

With the help of~\cite[Eqs.~(36)-(40)]{matrix_cookbook}, the notations in~(\ref{equ: gradient crlb wrt S-2}) can be derived as
\begin{align*}
& {\bm A_{i,j,u,v}^{\mathrm s}} = ([\bar{\bm \zeta}_{i,u}]_{(j-1)\gls{numFrame}+1:j\gls{numFrame}}\otimes \bm 1_{\gls{numFeed}}^\tran ) \\
& ~ \odot \! \gls{reshapeFunc}_{\gls{numFrame}\!\times\!\gls{numFeed}}\big(\big(([\dot{\bm H}_{i,v}^{\los}]_{(j-1)\gls{numFrame}+1:j\gls{numFrame}} \!\odot\! \bm C_i)[\odot] (\bm B_{i,j} \!\otimes\! \bm 1_{\gls{numFrame}}) \big) \bm 1_{\gls{numElem}}\big),
\end{align*}
\begin{align*}
{\bm B_{i,j,u,v}^{\mathrm s}} \!= \! &- \!([\bar{\bm \zeta}_{i,u}]_{(j\!-\!1)\gls{numFrame}\!+\!1:j\!\gls{numFrame}} \!\otimes\! \bm 1_{\gls{numFeed}}^\tran) \!\odot \!\gls{reshapeFunc}_{\gls{numFrame}\!\times\!\gls{numFeed}}\!\big(
\big(
[\bm K_{\mathrm{ft},i}]_{(j\!-\!1)\gls{numFrame}\!+\!1:j\!\gls{numFrame}}  \\
&[\odot] \big(
(\bm C_i [\odot] \bm B_{i,j} \!\otimes \!\bm 1_{\gls{numFrame}})(\bm V_i\bm T_i^{\hil})
\big)\big)
\bm\zeta_{i,v}
\big).
\end{align*}
Here, $\bm\zeta_{i,u} =\bm\varLambda_i^{-1}(\dot{\bm H}^{\los}_{i,u}\odot \bm T_i) \bm 1_{\gls{numElem}}\in\mathbb C^{\gls{numSBand}\gls{numFrame}\times 1}$, operator $[\odot]$ denotes the \emph{penetrating face product}, and function $\gls{reshapeFunc}_{\gls{numFrame}\!\times\!\gls{numFeed}}(\cdot)$ reshapes the vector in the argument to an $\gls{numFrame}\!\times\!\gls{numFeed}$ matrix.
Similarly, the corresponding notations for $\partial\gls{fimMat_userpos}/ \partial \gls{codeMat_i}$ can be obtained as
\begin{align*}
{\bm A_{i,u,v}^{\mathrm c}} = \sum_{j=1}^{\gls{numSBand}}  &([\bar{\bm \zeta}_{i,u}]_{(j-1)\gls{numFrame}+1:j\gls{numFrame}}\otimes \bm 1_{\gls{numElem}}^\tran) 
\!\odot \! [\dot{\bm H}^{\los}_{i,v}]_{(j-1)\gls{numFrame}+1:j\gls{numFrame}} \\
&\odot (\bm S_{i,j}\bm B_{i,j}),
\end{align*}
\vspace{-1.4em}
\begin{align*}
{\bm B_{i,u,v}^{\mathrm c}} = &- \sum_{j}^{\gls{numSBand}} \!
\big([\bar{\bm \zeta}_{i,u}]_{(j-1)\gls{numFrame}+1:j\gls{numFrame}}\!\otimes \!\bm 1_{\gls{numElem}}^\tran \! \odot\! (\bm S_{i,j} \bm B_{i,j})\big)\\
& \odot\! \gls{reshapeFunc}_{\gls{numFrame}\!\times\!\gls{numElem}}\!\Big(\!
[\bm K_{\mathrm{ft},i}]_{(j-1)\gls{numFrame}+1:j\gls{numFrame}}
[\odot]\!\big((\bm V_i\bm T_i^{\hil})\!\otimes\! \bm 1_{\gls{numFrame}}\!\big) \bm\zeta_{i,v}
\Big).
\end{align*}

\vspace{-1em}
\section{Proof of Proposition~\ref{prop: converge rate}}
\label{appx: proof of prop converge rate}

Based on~\cite[Lemma 2]{Ren2020Scheduling}, given parameter vectors denoted by $\bm a$ and $\bm b$, it can be derived that 
\beq
\label{appx equ: L ftl u}
\hat{\mathcal L}(\bm a) \leq \hat{\mathcal L}(\bm b) + \nabla_{\bm b} \hat{\mathcal L}(\bm b)^\tran(\bm a - \bm b) + \frac{L}{2}\|\bm a-\bm b\|^2_2.
\eeq
Substituting $\bm a =\bm w^{(t+1)} $, $\bm b = \bm w^{(t)}$, and $\nabla_{\bm b} \hat{\mathcal L}(\bm b)=\bm g^{(t)}$ into~(\ref{appx equ: L ftl u}), it can be shown that
\beq
\label{appx equ: L ftl w t+1}
\hat{\mathcal L}(\bm w^{(t+1)}) \leq \hat{\mathcal L}(\bm w^{(t)}) + (\bm g^{(t)})^{\tran}(-\gls{lrVec}\odot \hat{\bm g}^{(t)}) + \frac{L}{2}\| - \gls{lrVec} \odot \hat{\bm g}^{(t)}\|^2.
\eeq
where $\hat{\bm g}_x^{(t)} = -{Q_x}\Delta \bm w_x^{(t)}/({Q \xi^{(t)}_x})$.
Taking the expectation of both sides of (\ref{appx equ: L ftl w t+1}), we obtain 
\begin{align}
\label{appx equ: E L ftl w t+1}	
\mathbb{E} & \left(\hat{\mathcal L}(\bm{w}^{(t+1)})\right) \\
\leq & \mathbb{E}\!\left(\!\hat{\mathcal L}(\bm{w}^{(t)})\!\right)\!-\!\left(\bm{g}^{(t)}\right)^{\top}\! \gls{lrVec}\odot \mathbb{E}\!\left( \hat{\bm{g}}_x^{(t)}\right)\!+\!\frac{L}{2}\left( \gls{lrVec}\right)^{\circ 2 \tran} \!\mathbb{E}\!\left((\hat{\bm{g}}_x^{(t)})^{\circ 2}\right) \nonumber \\
= & \mathbb{E}\left(\hat{\mathcal L}(\bm{w}^{(t)})\right) - (\gls{lrVec})^\tran (\bm{g}^{(t)})\odot \mathbb{E}\left( \hat{\bm{g}}_x^{(t)}\right) \nonumber \\
&\qquad\qquad\qquad~ + \frac{L}{2}\left( \gls{lrVec}\right)^{\circ 2 \tran} \Big(\left(\mathbb{E}\big(\hat{\bm{g}}_x^{(t)}\big)\right)^{\circ 2}+\mathbb{V}\big(\hat{\bm{g}}_x^{(t)}\big)\Big) \nonumber \\
\stackrel{(a)}{=} &  \mathbb{E}\!\left(\hat{\mathcal L}(\bm{w}^{(t)})\right) \!-\!(\gls{lrVec})^{\tran}\!\Big( \bm 1\!-\!\frac{L}{2}\!\gls{lrVec}\Big)\!\odot\! (\bm{g}^{(t)})^{\circ 2}\!+\!\frac{L}{2}\!\left(\gls{lrVec}\right)^{\circ 2\tran} \!\mathbb{V}\!\left(\hat{\bm{g}}_x^{(t)}\right)\!,\nonumber
\end{align}
where~($a$) is because $\hat{\bm{g}}^{(t)}$ is an unbiased estimate of $\bm g^{(t)}$, and $\mathbb{V}\big(\hat{\bm{g}}^{(t)}_x\big) $ is the covariance of $\hat{\bm{g}}_x^{(t)}$:
\begin{align}
\label{appx equ: V}
\mathbb{V}\big(\hat{\bm{g}}_x^{(t)}\big) & \!=\! \mathbb E\big( (\hat{\bm g}_x^{(t)} \!-\! \bm g^{(t)})^{\circ 2}\big) = \mathbb E(\hat{\bm g}_x^{(t)\circ 2}) \!-\! {\bm g}^{(t)\circ 2}  \nonumber \\
& = \!\sum_{n=1}^{\gls{numUser}}\! \xi_n^{(t)}\!\Big(\! \Big(\frac{Q_n}{Q \xi_n^{(t)}}\!\Big)^2\!\mathbb E(\bm g_n^{(t)\circ 2}) \!+\! \Big(\frac{Q_n}{Q \xi_n^{(t)}}\!\Big)^2\!\mathbb E(\bm \varsigma_n^{(t)\circ 2}) \!\Big)\nonumber\\
& =  \sum_{n=1}^{\gls{numUser}} {1\over \xi_n^{(t)}} \cdot \Big(\frac{Q_n}{Q}\Big)^2 \cdot (\mathbb E(\bm g_n^{(t)\circ 2}) + \gls{dpSigma_n2}\bm 1).
\end{align}
Subtracting $ \mathbb{E}\big(\hat{\mathcal L}(\bm{w}^*)\big)$ from both sides of~(\ref{appx equ: E L ftl w t+1}), then (\ref{equ: converge rate}) in Proposition~\ref{prop: converge rate} is proven.

\vspace{-1em}
\end{appendices}

\vspace{1em}
\bibliographystyle{IEEEtran}
\bibliography{bibilio}

\end{document}